# Seasonal cycle of Precipitation over Major River Basins in South and Southeast Asia: A Review of the CMIP5 climate models data for present climate and future climate projections

Shabeh ul Hasson[1,2], Salvatore Pascale[2], Valerio Lucarini[2], Jürgen Böhner[1]

[1] CEN, Centre for Earth System Research and Sustainability, Institute for Geography, University of Hamburg, Hamburg, Germany

[2] CEN, Centre for Earth System Research and Sustainability, Meteorological Institute, University of Hamburg, Hamburg, Germany

*Correspondence to:* Shabeh ul Hasson   (shabeh.hasson@zmaw.de)
                                        (shabeh.hasson@uni-hamburg.de)

**Abstract**

We review the skill of thirty coupled climate models participating in the Coupled Model Intercomparison Project 5 (CMIP5) in terms of reproducing properties of the seasonal cycle of precipitation over the major river basins of South and Southeast Asia (Indus, Ganges, Brahmaputra and Mekong) for the historical period (1961-2000). We also present projected changes by these models by the end of century (2061-2100) under the extreme scenario RCP8.5. First, we assess their ability to reproduce observed timings of the monsoon onset and the rate of rapid fractional accumulation (RFA slope) - a measure of seasonality within the active monsoon period. Secondly, we apply a threshold-independent seasonality index (SI) – a multiplicative measure of precipitation (P) and extent of its concentration relative to the uniform distribution (relative entropy – RE). We apply SI distinctly for the whole monsoonal precipitation regime (MPR), westerly precipitation regime (WPR) and annual precipitation regime. For the present climate, neither any single model nor the multi-model mean performs best in all chosen metrics. Models show overall a modest skill in suggesting right timings of the monsoon onset while the RFA slope is generally underestimated. One third of the models fail to capture the monsoon signal over the Indus basin. Mostly, SI estimates for WPR are simulated higher than observed for all basins, while for MPR, it is simulated higher (lower) for the Ganges and Brahmaputra (Indus and Mekong) basins, following the pattern of overestimation (underestimation) of precipitation. However, models are biased positive (negative) for RE estimates over the Indus and Ganges (Brahmaputra and Mekong) basins, implying the extent of precipitation concentration for MPR and number of dry days within WPR higher (lower) than observed for these basins. Under the RCP8.5 scenario, most of the models project a slightly delayed monsoon onset, and a general increase in the RFA slope, precipitation and extent of its concentration (RE), all suggesting a higher seasonality of the future MPR for all basins. Similarly, a modest inter-model agreement suggests a less intermittent WPR associated with a general decrease in number of wet days and a decrease (increase) in precipitation over the Indus and Ganges (Brahmaputra and Mekong) basins. Based on SI, multi-model mean suggests an extension of the monsoonal domain westward over northwest India and Pakistan and northward over China. These findings have serious implications for the food and water security of the region in the future.

# 1 Introduction

Climate change has substantial impacts on the hydrological cycle at global (Allan, 2011; Kleidon and Renner, 2013; IPCC, 2013; Roderick et al., 2014), regional (Ramanathan et al., 2005; Lucarini et al., 2008; Turner and Annamalai, 2012; Hasson et al., 2013 and 2014a), and local scales (Roderick et al., 2014; Greve et al., 2014). The issue of understanding the impact of climate changes on the hydrological cycle has special relevance for areas dependent upon the seasonal water availability and for areas highly vulnerable to hydro-climatic extremes, such as South Asia (Hirabayashi et al. 2013; Hasson et al., 2013). Seasonal cycle of precipitation in South Asia is the key for ensuring food and water security of one-fifth of the world's inhabitants (Ho and Kang 1988; Lal et al., 2001; Sperber et al., 2013; Sabeerali et al., 2014) and strongly affects the gross domestic product of the agrarian-based economies (Subbiah et al., 2002; Gadgil and Gadgil, 2006). Historical observations show that the summer monsoonal rainfall over India has been decreasing significantly for the last six decades (Wapna et al., 2013; Bolasina et al., 2011) while the water table in most of India has already been dropped considerably (Rodell et al., 2009). The annual per capita water availability has substantially reduced and it still faces growing stress due to population growth and ongoing economic development (Babel and Wahid, 2008; Eriksson et al., 2009, Rasul, 2014). Such changes might become more pronounced under the global warming scenario with ensuing drastic impacts on the socio-economic setup in the region. Therefore, assessment of future changes in the precipitation regime and its seasonality are critical for the policy makers and relevant stakeholders for the future water resources management and long-term planning of the sustainable regional economies.

Presently, the global climate models are applied as a primary tool for projecting future changes associated with a variety of anthropogenic greenhouse gas (GHG) emission scenarios and are being extensively used to understand the climate system of the Earth and the hydrological cycle on global and regional scales. However, reliability of the projected changes largely depends upon the degree of skill of these models in adequately representing the physical climatic processes and in reproducing the observed hydro-climatic phenomena. Despite substantial improvements in their numerics and in the representation of the physical, chemical, and biological processes taking place in the climate system, a realistic representation of the hydrological cycle in these models has not been achieved, so far. This is due to the fact that in these models some of the crucial fine scale hydro-climatic processes that occur on a variety of spatio-temporal scales are not explicitly resolved but represented only through parameterization schemes (May, 2002; Hagemann et al., 2006; Tebaldi and Knutti, 2007). Furthermore, the structural limitations of the global climate models often lead to the underrepresentation of existent physio-geographical characteristics that greatly affect realism of the model simulations. Such inadequate representation may results in a serious bias in the crucial parameters and may lead to physical inconsistencies of water and energy balance at global (Lucarini and Ragone, 2011; Liepert and Previdi, 2012; Liepert and Lo, 2013) and regional scales (Lucarini et al., 2008; Hasson et al., 2013). Given these limitations, it is a great challenge for the present-day climate models to describe correctly the hydrological cycle over the South and Southeast Asia region that features a tremendous diversity in its hydro-climatic patterns, determined by its unique physio-geographical characteristics, mainly the extensive cryosphere and complex terrain of the Hindu Kush–Himalayan (HKH) ranges and Tibetan Plateau (TP). The hydrology of such region is determined by form and magnitude of the spatially heterogeneous and highly seasonal moisture input from the prevailing large scale circulations modes: the western (predominantly winter) mid-latitude disturbances (Wake, 1987; Rees and Collins, 2006; Ali et al., 2009) and the south and Southeast Asian summer monsoon, where the latter one dominates (Annamalai et al., 2007; Turner and Annamalai, 2012). The westerly disturbances are extratropical cyclones formed and/or fortified over the Caspian and the Mediterranean Seas, which transported through the southern flank of the Atlantic and Mediterranean storm tracks (Hodges et al., 2003; Bengtsson et al., 2006) to their far eastern extremity, enter into the study area along HKH and eventually subside over the continental India (Hasson et al., 2014a). On the other hand, the South and Southeast Asian monsoon along with the East Asian monsoon are interrelated components of the Asian monsoon system (Janowiak and Xie, 2003). The monsoon is a thermally driven system in which a large-scale meridional thermal gradient between the land and the ocean (Li and Yanai, 1996; Fasullo and Webster, 2003; Chou, 2003) is formed both, at the surface - due to intense seasonal solar heating over the land and its low heat capacity - and at the mid-to-upper troposphere - due to the HKH and TP causing sensible heating aloft forming the Tibetan warm anticyclone (Böhner, 2006; Clift and Plumb, 2008) - in late-spring that result in a north–south pressure gradient, which induces the cross-equatorial surface flow and heralds the monsoonal onset (Li and Yanai, 1996; Fasullo and Webster, 2003). The study basins of the Indus, Ganges and Brahmaputra are under the dominant influence of the south Asian summer monsoon precipitation regime (MPR), however, the Mekong basin receives its precipitation from both south Asian and southeast Asian components of the Asian monsoon system (Hasson et al., 2013). Over the study domain, the monsoon onset starts over the southern India and advances towards the southern China and subsequently to northwestern India and Pakistan (Matsumoto, 1997; Janowiak and Xie, 2003; Hasson et al., 2013 and 2014a). Hence, the monsoon onset starts over the Mekong and Brahmaputra basins in mid-to-late-May and tracks over the Ganges and Indus basins till June to July. The monsoon retreat, referring back to dry and dormant conditions, goes roughly in a reverse order, latest by October (Goswami, 1998). The sudden breaks during the active monsoon period (Ramaswamy, 1962; Miehe, 1990; Böhner, 2006), spanning over few days to several weeks (Turner and Annamalai, 2012), can seriously threat the water availability (Webster et al., 1998), and food production, particularly for the areas of rain fed agriculture (Subbiah, 2002; Gadgil and Gadgil, 2006), such as parts of the west India and Pakistan (Wani et al., 2009). These areas are also quite sensitive to the delays in the monsoon onset. Given that importance, an adequate model representation of the various local scale physio-geographical features and physical processes that are influential to the diversified aspects of the precipitation regimes associated with the prevailing large scale circulations, particularly the monsoon, are yet a great challenge for the climate science and modelling community.

It is unfortunate that despite the relative high-resolution, the current generation of climate models included in the 5$^{th}$ phase of the Coupled Model Intercomparison Project (CMIP5 - Taylor et al., 2012; Guilyardi et al., 2013) still misrepresent substantially the real topography of the HKH and Tibetan Plateau (Chakraborty et al., 2002 and 2006; Boos and Hurley, 2013) akin their predecessors. Moreover, the region features an extensive irrigation activity throughout the year that plays a major role in the interaction of the regional circulations over the region in order to determine the strength of the concurrent and subsequent monsoonal precipitation regimes and their spatial extents (Saeed et al., 2009 and 2013; Levine and Turner 2012; Marathayil et al., 2013; Levine et al., 2013). Such phenomenon is completely missing in the CMIP5 models. This further hinders a realistic simulation of the precipitation regimes over the region, particularly that of associated with the summer monsoon. As a result of such common deficiencies, the systematic errors in the simulated precipitation patterns, such as delayed monsoon onset over India, inadequate simulation of seasonal cycle (Kripalani et al., 2007; Kumar et al., 2011; Sperber et al., 2013; Sperber and Annamalai, 2014; Hasson et al., 2014a), underestimation of monsoonal precipitation and offset in the positions and intensities of its maxima (Wang et al., 2004; Annamalai et al., 2007;

Christensen et al., 2007; Lin et al., 2008; Sperber et al., 2013), rising trend of the monsoonal precipitation contrary to the observed falling trend (Wang et al., 2004; Ramanathan et al., 2005; Sabeerali et al., 2014; Saha et al., 2014), cold sea surface temperature (SST) biases over the northern Arabian Sea (Levine et al., 2013; Sandeep and Ajayamohan, 2014), and the suppression of the monsoon far north over China and far west over Pakistan are consistent between the CMIP 3 and 5 modeling efforts.

Despite of the above mentioned shortcomings, the CMIP5 models are successful in capturing some of the key features over the region with a high inter-model agreement. For instance, Sperber et al., (2013) have reported high fidelity of the CMIP5 models - and of their predecessors, included in the third phase of the Climate Models Intercomparison Project (CMIP3) - in simulating the mean monsoonal winds. Further, some studies (Menon et al., 2013; Sandeep and Ajayamohan, 2014) have shown that the coupled climate models are able to realistically simulate the recently observed northward shift of the low-level monsoonal jet, which is consistent with the future projected increase in the monsoonal extent (Kitoh et al., 2013; Lee and Wang, 2014) and overall widening of the tropical zone (Fu 2006; Seidel and Randel, 2007). Hasson et al. (2014a) have reported a satisfactory representation of the change in the western mid-latitude precipitation regime under climate change over the region by the CMIP3 models, featuring a qualitative agreement for the pole ward shift of mid-latitude storm tracks under the global warming scenario as documented by various studies (Bengtsson et al., 2006; Fu, 2006; Fu and Lin, 2011). Such capabilities of models - despite their structural limitations and incomplete representation of various features - provide some confidence on their projected changes.

As compared to the CMIP3, the CMIP5 models have gone through an extensive development, introducing higher horizontal and vertical resolutions, improved interactions between atmosphere, land use and vegetation components, interactive and indirect aerosols treatments and inclusion of a carbon cycle etc. (Taylor et al., 2012). The CMIP5 models show typically some improvements relative to the CMIP3 models in their ability to represent the climate system (Knutti and Sedláček, 2013; Knutti, 2013; Geil et al., 2013; Liu et al., 2014). For instance, future increase in the south Asian summer monsoon precipitation as projected by the CMIP3 models (Ashrit et al., 2003; Meehl and Arblaster, 2003; Ashrit et al., 2005; Hasson et al., 2014a) are still very uncertain due to a relatively larger inter-model spread (Meehl et al., 2007; Turner and Slingo, 2009; Kumar et al., 2011; Turner and Annamalai, 2012; Hasson et al., 2014a), which is related to the diversity in their spatial resolution (Kim et al., 2008) and in the adopted convection schemes (Turner and Slingo, 2009; Bollasina and Ming, 2012). Contrary to this, the CMIP5 models feature a high agreement on the intensification of the summer monsoon precipitation (Kitoh et al., 2013; Menon et al., 2013; Wang et al., 2013; Lee and Wang, 2014) in future. Moreover, the CMIP5 models show a relative improvement for the time-mean precipitation through better simulating its maxima over the region of steep topography (Sperber et al., 2013), which may be linked to their relatively fine resolutions (Rajendran et al., 2013).

Most of the studies on the seasonal cycle of precipitation over South Asia focus only on MPR and adopt various threshold-based techniques to investigate fidelity of the climate models in representing, e.g., timing of the monsoon onset, retreat and maxima (e.g. Sperber et al., 2013; Sperber and Annamalai 2014; Hasson et al., 2014a). Since the westerly disturbances intermittently transport moisture to the region (Syed et al., 2006; Hasson et al., 2014a), such metrics are not equally applicable for the westerly precipitation regime (WPR). Moreover, an absolute threshold of 5 mm day$^{-1}$ is commonly used to identify the monsoon onset and retreat (Wang and LinHo, 2002; Sperber et al., 2013), which in view of the large inter-model discrepancies does not seem appropriate for testing models performances, especially when large biases in the seasonal cycle of precipitation are present (Geil et al., 2013). Namely, a dry (wet) bias may lead to a delayed (early) timing of the onset relative to the observations. Some of the dry models even cannot achieve the absolute threshold, particularly over the arid regions and areas of far reaches of the monsoon where amplitude of the simulated precipitation can be very small, such as, over the Indus Basin (Hasson et al., 2013 and 2014a). Therefore, in order to define the monsoon onset and retreat from a large inter-model "wetness" scatter the use of a relative threshold has been encouraged (Geil et al., 2013; Hasson et al. 2014a; Sperber and Annamalai, 2014). Analyzing the CMIP3 models, Hasson et al. (2014a) have uniformly applied a relative threshold of 0.17 over the normalized basin-integrated monthly precipitation of the Indus, Ganges, Brahmaputra and Mekong basins. Using a relative rather than an absolute criterion of accumulated precipitation has considerably reduced the effect of a large CMIP3 inter-model spread on defining the monsoon onset as opposed to typical threshold-based techniques. Similarly, Sperber and Annamalai (2014) have applied a relative threshold of 0.2 over the fractional accumulated pentad (5-day mean) precipitation from the CMIP5 models. Though use of a relative threshold together with the adopted data processing approach circumvents the effect of a large inter-model scatter on the analysis (Hasson et al., 2014a; Sperber and Annamalai, 2014), but again, uniformly applying a relative threshold over the whole study domain may face difficulties because of the spatial heterogeneity of the precipitation regimes and/or the temporal data resolution used in the analysis.

The goal of this review is to provide a comprehensive analysis of the performance of CMIP5 models in representing the seasonal cycle of precipitation and assessing their suggested changes in response to changing climate conditions over the major river basins of South and Southeast Asia, namely, Indus, Ganges, Brahmaputra and Mekong. Unlike previous studies which assess models' realism either at the grid-point or at the regional scale (Annalmali et al., 2007; Fan et al., 2012; Turner and Annamalai, 2012; Sperber et al., 2013; Menon et al., 2013), the focus of this study is on a river basin scale (Lucarini et al., 2008; Hasson et al., 2013, 2014a). A river basin is a natural unit of practical water resource management that is also quite relevant for impact assessment studies and can support for taking effective adaptive measures (Lucarini et al., 2008; Hasson et al., 2013 and 2014a). First, we assess models' performance for the time-dependent characteristics of MPR that are crucial for the water resources management and subsequently for the food security in the region. In this regard, we choose timing of the onset and retreat and the rate of precipitation concentration during the wet season as skill metrics (monsoonal metrics) for twofold reason; 1) these are stringent tests for assessing the model performance and, 2) these are most commonly employed by various model performance assessment studies, so our results from the latest generation climate models are comparable (e.g. Hasson et al., 2014a; Sperber and Annamalai, 2014). In order to circumvent a large inter-model scatter of the CMIP5 models, we perform a relative threshold based assessment of model performance over each basin based on the fractional accumulated precipitation (Sperber and Annamalai, 2014). First, we identify a unique set of relative thresholds for each basin from the observational dataset used in the study that yields right timings of the monsoon onset and retreat for the respective basin. Then, we apply these thresholds to the model datasets in order to investigate their skill for the select metrics and to assess their suggested changes. Such basin-scale evolution of monsoon only precipitation regime is, however, limited to the active monsoon duration (onset-retreat) and highly sensitive to its accurate identification.

Therefore, we additionally apply a novel and threshold independent technique over both MPR and WPR in order to distinctly characterize their overall seasonality over the study basins as well as over the spatial domain of South and Southeast Asia (Fig. 1). Such technique combines use of mean precipitation (P) and recently introduced dimensionless seasonality index (SI). The SI, accounting for the spatio-temporal heterogeneity of precipitation distribution, identifies the high-seasonality hotspots. The SI for a considered time period is determined by multiplicatively combining the precipitation (P) normalized by its spatial maximum and the relative entropy (RE) – an information theory - based quantitative measure of the extent of precipitation concentration against the uniform distribution. The index was first employed by Feng et al. (2013) using monthly observations for the tropical land regions and then by Pascale et al. (2014a,b) over the monthly precipitation from the CMIP5 models for the whole globe.

Since changes in the time behavior of the south Asian monsoon precipitation regime are not evident at monthly scale (Hasson et al., 2014a), here, we investigate both monsoonal metrics and seasonality indicators on pentad precipitation datasets. In the first part of study, we report general and consistent biases of the CMIP5 models for all the chosen metrics against the observations during the historical period (1961-2000). This provides useful information for the climate modelers, as they suggest possible ways to improve the climate models, as well as to the practitioners and agencies working with the limited area models in the region of interest, as biases of individual models are made clear. In the second part, we present projected future changes in the select metrics by the end of the century (2061-2100) under an extreme representative concentration pathway (RCP) RCP8.5. The RCP8.5 scenario, referring to highest GHG emissions due to burgeoning population, slow income growth, modest technological development, higher energy demands and absence of climate change policies (Riahi et al., 2011), provides a reasonable upper bound to climate change, and constitutes a plausible worst-case-scenario, so studying it is important for bracketing the future possible impacts of climate change in the hydro-climatology of the region.

The paper is organized as follows. In section 2, we briefly describe the data used in the study and the methodology adopted for the analysis. Section 3 presents the model performance and their projected changes for the chosen metrics. In section 4, we discuss in detail the findings from the present study in relation with the available literature and summarize the main results.

## 2 Data and method

It is indeed nontrivial to reconstruct the statistics of gridded datasets of precipitation in South and Southeast Asia (Collins et al., 2013). The main problems come from the observational uncertainty typically associated to: 1) an uneven density and temporal coverage of in-situ observatories; 2) performance of the applied interpolation techniques; 3) difficulties in snow detection over the complex HKH and Tibetan Plateau terrain (Fekete et al., 2004; Yatagai et al., 2012; Palazzi et al., 2013; Hasson et al., 2014a; Prakash et al., 2014). Therefore, we have decided to use multi-source satellite-merged observations for our analysis. We have used the pentad precipitation datasets from the Global Precipitation Climatology Project (GPCP, Xie et al., 2003) and the Climate Prediction Center Merged Analysis of Precipitation (CMAP, Xie and Arkin 1997a) for the period 1979-2004 that were available at $2.5^{\circ} \times 2.5^{\circ}$ horizontal grid resolution. The GPCP precipitation was obtained through merging microwave and infrared based observations with in-situ rain gauge data, while the CMAP precipitation was obtained through merging the microwave, infrared and outgoing long-wave radiation based observations with the NCEP-NCAR reanalysis dataset (Huffman et al., 1997 and 2009; Xie and Arkin, 1997b). Various studies have used the GPCP and CMAP datasets for the global (Martin and Levine, 2012; Kitoh et al. 2013; Chou et al. 2013; Frierson et al. 2013; Ramesh and Goswami, 2014) and regional (Bolvin et al., 2009; Cook and Seager 2013; Sperber et al. 2013) scale analysis of the hydrological cycle. Sperber and Annamalai (2014) have also shown that the two datasets show little differences in terms of time dependent properties of the seasonal cycle of precipitation over the south Asian summer monsoon domain. Xie et al. (2003) have documented the development of pentad GPCP dataset and discussed its differences relative to the CMAP observations. As of the model simulations, daily precipitation data were obtained from the CMIP5 data for the historical period (1961–2000) and for the projections (2061–2100) under an extreme global warming scenario of RCP8.5 (Table 1). The RCP8.5 assumes that the radiative forcing will ramp up to 8.5Wm$^{-2}$ by the year 2100 and can be considered the upper end of the climate forcing scenarios adopted by the IPCC. Further details about the RCPs can be found in Moss et al. (2010) and van Vuuren et al. (2011). The requisite data was available from thirty coupled climate and Earth System Models (ESMs) that provide several ensemble members for each simulated scenario. We have chosen only first realization from each model as Coleman et al. (2011) and Pascale et al. (2014a & b) have reported robustness among the several ensemble members of an individual model.

For the monsoonal metrics, we have first calculated the time series of basin-integrated pentad precipitation from each considered dataset for the four study basins. For each year and for each pentad, the basin-integrated precipitation was calculated through first weighing the pentad precipitation at each grid cell by the fraction of its area lying within the natural boundary of a basin and then area-averaging over the basin area:

$\langle p \rangle_{i,k} = \frac{1}{A} \int_A p_{i,k} \, dA$        Eqn. 1

where $p_{i,k}$ is the mean precipitation at pentad $i$ and year $k$ at a certain grid point, $A$ is the basin area and $\langle p \rangle_{i,k}$ is the basin-averaged precipitation at the same time. Further details about the basin integration approach for precisely calculating volumetric quantities can be found in Lucarini et al. (2008) and Hasson et al. (2013 and 2014a). We have then accumulated $\langle p \rangle_{i,k}$ (Eqn. 2) such as each pentad $i$ is the sum of its preceding pentads within a year $k$ and itself (in the following we omit the label $k$ to keep the notation as simple as possible) and from the accumulated precipitation, $\Pi_t$, we estimate the fractional accumulated precipitation, $\widetilde{\Pi}_t$:

$\Pi_t = \sum_{i=1}^{t} \langle p \rangle_i$        Eqn. 2

$\widetilde{\Pi}_t = \Pi_t / \Pi_{73}$        Eqn. 3

where $t = 1, 2, 3, \ldots, 73$ denotes each pentad of the year, as models with Gregorian and 360 days calendar were adjusted to 365-day calendar. We plot seasonal cycles of $\Pi_t$ from all datasets in Figure 3 to show the inter-model scatter of the simulated precipitation, multi-model mean (MMM) and observations for each basin. In order to circumvent the effect of such scatter on the estimation of select metrics, we have obtained the fractional accumulation, $\widetilde{\Pi}_t$ by dividing each pentad of $\Pi_t$ within a year by its last pentad (Eqn. 3). Based on $\widetilde{\Pi}_t$ of the GPCP/CMAP datasets, we have identified relative thresholds for the observed timings of monsoon onset and retreat for each basin (see Fig. 4). We have considered a threshold for the monsoon onset as the fractional accumulation of the pentad after which the rapid fractional accumulation (RFA) starts, $\widetilde{\Pi}_{on}$. Similarly, for the monsoon retreat, it was taken as the fractional accumulation of the pentad, before which the rapid fractional accumulation ends, $\widetilde{\Pi}_{ret}$. The RFA refers to the linear growth in the fractional accumulation within the active monsoon duration and can be determined through a linear slope (onward called RFA slope) calculated using Eqn. 4.

$RFA\ Slope = S = (\widetilde{\Pi}_{ret} - \widetilde{\Pi}_{on})/N$       Eqn. 4

where $N$ refers to the total number of pentads during the active monsoon duration (from $\widetilde{\Pi}_{on}$ to $\widetilde{\Pi}_{ret}$). The RFA slope tells how fast the precipitation accumulates within an active monsoon period (Fig. 5). In other words, it provides a measure of the concentration of precipitation within the active monsoon duration. We have verified that our identified thresholds for each basin, based on the GPCP/CMAP datasets, provide right timings of the climatic monsoon onsets and retreats for the respective basins when compared to the observed onset/retreat isochrones - lines spatially indicating progression of the long term mean monsoon onset/retreat dates - around the center of the respective basins as defined by the Indian Meteorological Department (Krishnamurti et al., 2012) and as identified by Janowiak and Xie (2003) from the GPCP dataset. Based on these thresholds, we have first obtained for each basin the timing of the monsoon onset, retreat and the RFA slope on yearly basis and then their climatological means were obtained for the historical (1961-2000) and future (2061-2100) periods. We estimated the offsets of the models from the observations during the historical period, and assessed the future projected changes with respect to historical period for the considered metrics along with their statistical significance using a Students' T-test (Welch, 1938).

Since the study area undergoes the influence of two large-scale circulation modes, the summer monsoon and westerly disturbances, some regions feature bimodal precipitation regime, yielding substantial amounts of water in distinct periods of the year (Hasson et al., 2014a). The central and eastern part of the region receives water almost exclusively in summer, during the monsoon (MPR). In addition to the monsoonal precipitation, the western part of the study area receives water during winter and spring seasons (Rees and Collins, 2006; Hasson et al., 2013 and 2014a) as a result of the synoptic disturbances from the southern flank of the jet (WPR). In view of such spatial heterogeneity, we have analyzed the spatial dependence of the SI, in addition to analyzing it on the aggregated basin scale. This has allowed us to present local and sub-basin scale model performance and their projected changes. In order to apply SI separately over both MPR and WPR, we have roughly divided the hydrological year into monsoon and "westerly" seasons, based on our results from the fractional accumulation analysis. We refer the period May to October as the MPR (36 pentads), the rest of the hydrological year (November to April) is taken as WPR, comprising of 37 pentads (See Fig. 2).

Since performance of the models in our above mentioned analysis highly depends upon the accuracy of the chosen relative thresholds, and thus on the adequate identification of the active monsoon duration, we have additionally employed threshold-independent seasonality index (SI) (Feng et al., 2013; Pascale et al., 2014a, b), which, accounting for the spatio-temporal distribution of precipitation, provides a measure of the overall seasonality of precipitation regime. The SI is given by the product of the precipitation P (normalized, for convenience, by its spatial maximum) times the relative entropy RE (Eqn. 5) – an information theory based quantitative measure of the degree of concentration of the precipitation, measuring the "distance" from the uniform distribution. Considering a grid cell $x$ in our domain and a specific time range of WPR, MPR and annual precipitation regime comprising of $t$ pentads, let us define $\pi$ as the precipitation fraction of the pentad $i$:

$$\pi_{i,x} = p_{i,x}/P_x \qquad \text{where } i = 1...t \text{ and } P_x = \sum_{i=1}^{t} p_{i,x} \qquad \text{... Eqn. 5}$$

from which the relative entropy (*RE)* is computed as:

$$RE_{t,x} = \sum_{i=1}^{t} \pi_{i,x} \log_2 (t\, \pi_{i,x}) \qquad \text{Eqn. 6}$$

where *t* is 73 for the hydrological year, 36 for the monsoon season and 37 for the cold season. Note that when applying the RE indicator to the MPR or WPR, we are indeed looking at the subseasonal variability of precipitation.

The RE is maximum ($= \log_2 t$) when the precipitation is concentrated in a single pentad, and zero when the precipitation is uniformly distributed among the pentads ($\pi_i = 1/t,\ i = 1...t$) within the considered time duration. Changes in RE is, therefore, refer to the effective dryness or wetness of the pentads within the considered time period. The RE changes, purely caused by the transformation of pentads from wet to completely dry and vice versa, can also be related to changes in the length of dry/wet season in the case of MPR. The RE is related to number of dry days, $\tilde{n}$ (Pascale et al., 2014a):

$$\tilde{n} = 5t(1 - 2^{-RE}) \qquad \text{Eqn. 7}$$

where *t* refers to number of pentads in a considered precipitation regime. Here, an increase (decrease) in RE estimates corresponds well to the increase in number of dry (wet) days. Moreover, SI of WPR when dominated by RE estimates need to be interpreted better as an indicator of erratic behavior, since P comes from number of incident events. Based on estimates of RE and P, SI for all considered time periods is obtained as Eqn. 8.

$$SI_{x,t} = RE_{x,t} \frac{P_{x,t}}{P_0} \qquad \text{Eqn. 8}$$

Here $P_0$ is a constant scaling factor and it is taken as the maximum of $P_{x,t}$ within the considered time period, *t* of the observed datasets GPCP/CMAP (Feng et al., 2013). The estimate of SI is zero if either *RE* or $P_{x,t}$ is zero (completely wet or dry conditions, respectively) and reaches its maximum value when *RE* estimate for the year of $P_0$ approaches maximum ($\log_2 t$). The SI index provides a measure of seasonality by considering the spatial heterogeneity of the precipitation distribution. For instance, over the Thar Desert (on the southern portion of the Pakistan-India border), SI will be low despite of the erratic behavior of its precipitation and prolonged dry season (highest RE) because the incident precipitation is extremely low. On the other hand, regions under year round precipitation regime (low RE) such as eastern Tibet can still feature high SI because total precipitation input is high. Details of the seasonality indices and their calculation are given in Feng et al. (2013) and Pascale et al. (2014a, b). In order to consistently estimate *RE* and *SI*, we have remapped all model datasets to the common GPCP and CMAP grid resolution of 2.5° x 2.5° using a bilinear interpolation. For sake of inter-model comparison, we have adjusted the model datasets with the Gregorian and 360-day model calendar (Table 1) to 365-day calendar before the pentad precipitation was calculated. We have calculated the yearly estimates of seasonality indicators (P, RE, SI) from the gridded pentad precipitation, integrated them over the study basins (Eqn. 1) and obtained their climatological means for the historical (1961-2000) and future (2061-2100) periods.

It is worth mentioning that in the rest of the paper we do not perform bias reduction/correction of the data, in order to have the ability to fully appreciate the models' discrepancies. Furthermore, our presented results are mainly relative to the GPCP dataset as it performs relatively better than the CMAP dataset in terms of the monsoonal precipitation (Prakash et al., 2014). In fact most of the available observations feature certain limitations in the rainfall estimation over the mountainous massifs, along the Himalayan foothills and over the northeast India (Prakash et al., 2014). Thus, we suggest that our results should be interpreted in the context of a larger spectrum of observational uncertainty featured by both the observational datasets considered here, and those discussed elsewhere (e.g. Collins et al., 2013; Prakash et al., 2014).

## 3 Results

### 3.1 Monsoon onset, duration and RFA slope (1961-2000)

The GPCP data set agrees well with the CMAP observations for the climatological seasonal cycle of basin-integrated P over the study basins (Fig. 2 and 3). Instead, models exhibit diverse skills, featuring a large inter-model spread and deviation from the observations. It is clear from the figure 3 that models perform well over the Mekong basin in absolute term while models overestimate P for the Brahmaputra basin and show both overestimation/underestimation of P for the Indus and Ganges basins. The $\Pi_t$ plotted in figure 3 for the Indus basin clearly shows that some models do not adequately simulate the MPR (no sharp growth between pentads 37-52). In figure 4, we show that $\widetilde{\Pi}$ eliminates the inter-model scatter of P (shown in Figs. 2 and 3) and that in relative terms models reasonably simulate the seasonal cycles for the Mekong, Brahmaputra and Ganges basin, which are dominated by the MPR. Here, a sharp growth in $\widetilde{\Pi}_t$ shown in figure 4 refers to the active monsoon duration that spans over 26-62, 28-57, 32-55 and 37-52 pentads, for the Mekong, Brahmaputra, Ganges and Indus basins, respectively. Interestingly, we note that models for the Indus basin feature serious discrepancies and are generally biased wet for the WPR but dry for the MPR as compared to the considered observations. For instance, the ratio of observed precipitation between the monsoon period and rest of the year is around 1.46, where the models range between 0.18-1.34, where only six models feature such ratio above unity. In figure 5, we plot the estimated RFA slope against the identified monsoon onset pentads from all datasets, which clearly summarizes adequacy of the individual models and of MMM for all basins, against the observations.

As discussed, the Indus basin features a bimodal precipitation distribution because of the contributions of the westerly disturbances and summer monsoon system (Syed et al., 2006; Ali et al., 2009; Palazzi et al., 2013; Hasson et al., 2014a). We note that most of the models during such period (pentads 1-36) overestimate $\widetilde{\Pi}$ against the GPCP/CMAP observations. For the monsoon season, we have found fractional accumulation of 0.15 and 0.9 as relative thresholds that identify right timings of the monsoon onset and withdrawal, respectively (see Fig. 4). Thus, the monsoon duration spans over the pentads from 37 to 52 comprising a total of 16 pentads (Fig. 4). During such period, the GPCP data set suggests a highest RFA slope of 0.048 for the basin, indicating shortest active monsoon duration among all study basins. In order to eliminate the effect of WPR while analyzing the monsoonal properties, here, we consider only pentads from 31-73. The difference between the CMAP and GPCP datasets in terms of their suggested climatic properties of the onset, duration and RFA slope for the basin is statistically insignificant. The MMM suggests a realistic timing of the monsoon onset, it prolongs the monsoon duration by 10 pentads and underestimates the RFA slope by 45% (Table 2). Eight models (CanESM2, CESM1-CAM5, CSIRO-Mk3-6-0, IPSL-CMA-MR, MIROC5, MPI-ESM-LR and MRI models) suggest realistic timings of the onset (Fig. 5). The rest of models generally suggest a delayed onset by 3-8 pentads, except the MIROC-ESM, MIROC-ESM-CHEM and models from the MOHC. Irrespective of the models' skill regarding the onset, all models suggest an unrealistically long duration of monsoon by 5-15 pentads with respect to observations. We note that such prolonged duration of the monsoon is linked with the substantial underestimation of the RFA slope (or vice versa) for the basin (see Fig.5 and Table 2). Actually, both estimates are highly negatively correlated. Interestingly, in one-third of the models (bcc-csm1-1, IPSL-CMA-LR, IPSL-CMB-LR, CSIRO-Mk3-6-0, FGOALS-g2, INMCM4 and models from CMCC and MRI) MPR is almost absent over the basin. Two other models (EC-EARTH and BNU-ESM) show an unrealistic seasonal cycle of precipitation.

For the Ganges Basin, our identified relative thresholds for timings of the onset and withdrawal of the monsoon are pentads of 0.15 and 0.95 fractional accumulation, respectively. The observed climatic monsoon onset starts at the pentad 32 while active duration of the monsoon spans over 25 pentads. The RFA slope is estimated to be 0.036. We note that two models (IPSL-CMB-LR and EC-EARTH) simulate overall an unrealistic seasonal cycle of P for the basin. These models along with seven other models (bcc-csm1-1, FGOALS-g2, IPSL-CMB-LR and models from MOHC and MRI) achieve quite early the onset threshold (before pentad 26) but their rapid accumulation starts quite later (Fig. 5, Table 2). This may be attributed to the unrealistic pre-monsoonal P or to the prolonged WPR simulated by these models. Such a systematic wrong onset influence the MMM, which also suggests an unrealistically early onset, with a bias of four pentads. Most of the models suggesting an unrealistically early onset mainly overestimate the observed duration of the monsoon and substantially underestimate the rapid accumulation during such period. From the rest, 17 models (BNU-ESM, CanESM2, CMCC-CMS, CMCC-CM, CNRM-CM5, CSIRO-Mk3-6-0, INMCM4, IPSL-CMA-MR, NorESM1-M and models from MPI, MIROC, NCAR and NSF-DOE-NCAR) suggest right timings of the onset. Only models from NOAA-GFDL suggest a statistically significant delay in the onset timing by 2-4 pentads. Six models (CESM1-CAM5, GFDL-ESM2G and GFDL-ESM2M, MIROC5 and models from MPI) suggest realistic duration of the monsoon.

For the Brahmaputra basin, the pentads suggesting right timings of the monsoon onset and withdrawal are those featuring fractional accumulation of 0.18 and 0.95, respectively. The active monsoon regime starts at pentad 29 and ends at pentad 57, spanning over 28 pentads in total. The RFA slope is about 0.030, which is higher than the Mekong basin but lower than the Ganges and Indus basins. The CMAP dataset suggests an early onset by one pentad and a prolonged duration by 2 pentads while it underestimates the RFA slope by 5% against the GPCP dataset. In view of the observational uncertainty, half of the models suggest realistic timings of the onset (See Table 2). Two models (BNU-ESM and NorESM1-M) suggest an onset delayed by 2 pentads, while the rest of 13 models suggest early onset by 2-11 pentads. Two models (EC-EARTH and IPSL-CMB-LR) perform worst in terms of all monsoonal metrics. For the RFA slope, only two models (MPI-ESM-MR and CSIRO-Mk3-6-0) agree with the observations while all the rest of models underestimate it (Fig. 5 and Table 2). Such models also simulate extended monsoon duration, as the RFA slope is highly negatively correlated to the monsoon duration. The MMM suggests an early onset by 2 pentads and a duration prolonged by 7 pentads while the RFA slope is underestimated by 16% as compared to the GPCP dataset. An early onset suggested by the MMM is mainly influenced by two models (EC-EARTH and IPSL-CMB-LR) performing worst in terms of the onset timings (Fig. 5).

The Mekong basin features a gentle RFA slope (0.025), an early onset (at pentad 26) and extended active monsoon duration (36 pentads) among all study basins. The fractional accumulation of 0.11 and 0.95 suggest pentads of the monsoon onset and withdrawal for the basin. The difference between CMAP and GPCP datasets in terms of the chosen metrics is insignificant. The MMM also suggests a realistic timing of the onset and duration while it overestimates the RFA slope by only 5%. Only seven models (bcc-csm1-1, BNU-ESM, CMCC-CESM, CMCC-CMS, FGOALS-g2, GFDL-ESM-2G, NorESM1-M and MPI models) suggest realistic timings of the onset. Thirteen models suggest the onset timing delayed while nine models suggest it earlier by 1-4 pentads (Fig. 5 and Table 2). Surprisingly, EC-EARTH model suggests timing of the onset earlier by 20 pentads. Such a model performs consistently worst for all metrics over the Mekong and over all other study basins. Ten models suggest a realistic duration of the monsoon suggesting either smallest or no difference in their estimated RFA slopes against the GPCP observation. Thirteen models suggest the monsoon duration shortened by 2-7 pentads while only five models suggest it extended by 2-8 pentads. Note that Mekong is the only study basin for which the monsoon duration is simulated significantly shorter than the observed by almost half of the models, mainly because of delayed onset (Fig. 5).

## 3.2 Mean precipitation, relative entropy, and seasonality index (1961-2000)

In addition to presenting basin-integrated estimates of seasonality indicators (Table 3), we discuss in detail the local scale biases from the individual models for such indicators. Here, we show estimates of P, RE and SI (Figs. 6-8, respectively) for the GPCP and CMAP observational datasets and for the MMM (top to bottom rows, respectively) for WPR, MPR and annual precipitation regime (left to right columns, respectively). Figures showing biases and projections from the individual models are given in the supplementary material. Table 4 summarizes the root mean square error (RMSE), pattern correlation (PC) and standard deviation ($\sigma$) of P, RE and SI calculated over the whole spatial domain (as in Fig. 1), for the observations, MMM and for all individual models within WPR and MPR. We remind here that RE estimates for WPR are more appropriate to be interpreted as number of dry/wet days due to the fact that westerly P is received over the region from number of intermittent events. Thus, for such areas, we interpret SI as an indicator of intermittency or erratic behavior.

**Westerly Precipitation Regime**

For all study basins, most of the models suggest higher values of SI during WPR as compared to the observations. This is generally associated with overestimation of P, irrespective of the fact that number of dry days (RE estimates) are overestimated (underestimated) in the Indus and Ganges (Brahmaputra and Mekong) basins with respect to the observations.

Looking at spatial variability of these estimators, one finds that the biases in the P, RE and SI are not uniform among the individual models (See supplements). As of P, most of the models considerably overestimate it over the HKH region, with the exception of eight models (EC-EARTH, CSIRO-Mk3-6-0, GFDL-ESM2M, GFDL-ESM2G, MRI-CGCM3, CanESM2,

IPSL-CMA-MR and CMCC-CESM). Models from the MIROC, NCAR and NSF-DOE-NCAR, along with five models (FGOALS-g2, bcc-csm1-1, BNU-ESM, EC-EARTH and NorESM1-M) slightly overestimate P over India, where the rest of models slightly underestimate it. Most of the models underestimate P over southern Pakistan (lower Indus Basin). The estimates of RE are substantially higher than observed over most of Pakistan, India, Myanmar and Thailand by the products of the CMCC, MRI, MPI, IPSL, and NOAA-GFDL modeling groups, suggesting higher number of dry days within WPR with respect to the GPCP observations. The rest of models mostly show lower than observed RE over southern India and Tibetan Plateau, implying higher number of wet days within WPR therein. Positive biases in SI are observed over most of the target region by almost all models, and particularly over the HKH ranges. Thus, the MMM also provides value for SI larger than observations over the HKH ranges, Tibetan Plateau and eastern China regions, due to overestimation of P. The GPCP, CMAP and MMM agree well on the extremely high estimates of RE over the Thar Desert and over the adjacent (very dry) areas of Pakistan and India, suggesting minimum number of wet days therein. In contrast, MMM overestimates the observed number of wet days over the Myanmar coast and Taklimakan Desert (north of the Tibetan Plateau), where low values of RE (dry days) are found. Over the Brahmaputra region, medium level estimates of P, RE and SI are observed. Fig. 6 clearly shows that northern Pakistan (northern Indus basin) and Brahmaputra basins receive considerable P under westerly disturbances.

**Monsoonal Precipitation Regime**

For the study basins, more than half of the models have a positive (negative) bias for RE for the Indus and Ganges (Brahmaputra and Mekong) basins. However, in contrast to WPR, models feature negative (positive) bias for the SI for the Indus and Mekong (Ganges and Brahmaputra) basins, mainly due to the underestimation (overestimation) of P.

We now focus on the spatial variability of the indicators. The GPCP dataset features high estimates of P over the Western Ghats of India, northern India (Ganges Basin), Bangladesh (and the whole Brahmaputra Basin), Myanmar, and over the Mekong Basin. The CMAP dataset agrees well with the GPCP dataset in terms of spatial correlation (PC=0.85), whereas the RMSE is around 300mm. Such difference is mainly due to the higher P over the Bay of Bengal in the CMAP dataset and over Myanmar and Brahmaputra Basin in the GPCP dataset relative to each other. Both observational datasets suggest a negligible P over the southern Pakistan (lower Indus Basin region), the western Tibetan Plateau and over the Karakoram Range. In contrast, MMM underestimates P over the northwestern India while it hardly sees any precipitation over the Indus basin. Such systematic biases mainly arise due to well-known inability of most of the models in fetching the monsoonal system too far to its western extremity. Similarly, MMM overestimates P over the eastern parts of the Himalayas and Tibetan Plateau region. Besides of the bias in P, low values of RE found over the lower Mekong Basin, over the Brahmaputra basin and over the wide area of the eastern Tibetan plateau are somewhat consistent between the GPCP, CMAP and MMM datasets. We find a negative bias in SI over the Indus basin and over large parts of the western Tibetan Plateau, which is associated with large RE and low P estimates. On the other hand, low estimates of SI over the lower Mekong Basin and eastern Tibetan Plateau are associated with the presence of a too regular regime of precipitations (low RE). We note that the RE increases as one moves northwest, thus implying a more concentrated MPR in that direction. Such transition is however not fully consistent between the observations and the MMM, as the latter features a positive bias in RE over large part of northwestern India. It is also worth noting that the region defined by values of SI below 0.11 for the MPR corresponds fairly well to the actual spatial extent of the south Asian summer monsoon. This also matches well with the domain estimated by Pascale et al. (2014a) using SI ≤ 0.05 and Wang et al. (2011) who used a threshold of 2.5 mm day$^{-1}$ over the estimated annual precipitation range for the study region. Such consistent results show that the extent of the monsoon can also be defined not only by resorting to an index capturing a seasonal variability, but also considering indicators of sub-seasonal variability, as in case of present study. The MMM suggested monsoon domain is underestimated over northwest India, Pakistan and over China regions against the observations, which also feature small differences among each other.

In contrast to the low bias of P of MMM over northern India, models from the NCAR, NSF-DOE-NCAR and NOAA-GFDL as well as the model EC-EARTH either suggest negative or no biases against the observations. The rest of the models suggest higher bias either over the whole India or at least over the Ganges Basin (northern India). Models from NCAR, NSF-DOE-NCAR and MPI along with two models (MIROC5, NorESM1-M) suggest very large positive bias of P, and SI over the Himalayan range and almost no bias for RE. Models from the MRI along with two models (IPSL-CM5B-LR and CSIRO-Mk3-6-0) show over the whole monsoonal domain the largest dry bias, with highest RE and lowest SI. Data from the MPI, IPSL, NOAA-GFDL, CMCC, MOHC and BCC modeling groups suggest higher RE mainly over Pakistan (Indus Basin). The remaining (almost half) of the models generally underestimate RE over the whole domain. A general underestimation of SI by most of the models is either associated with the underestimation of RE or P, or both.

**Annual precipitation regime**

Based on the analysis of basin-integrated seasonality indicators, we note from most of the models that seasonality of the annual precipitation regime over the Indus (Mekong) basin is generally dominated by SI estimates of WPR (MPR) (Table 2). For the mean annual P, the GPCP and CMAP datasets substantially differ - CMAP suggests higher magnitude of P particularly over the Western Ghats of India and over the lower Mekong Basin (coast of Cambodia), while the GPCP suggests the same over the Brahmaputra basin (mainly over Bangladesh, Myanmar and Indian province of Assam). Such differences mainly arise from the multi-probed data sources merged in these datasets. There is a negligible P over the Thar Desert, Taklimakan Desert and over the southern Pakistan, as also shown by MMM. All three datasets, therefore, indicate highest estimates of RE and lowest estimates of SI, suggesting a low number of wet days and highly erratic precipitation regime over these regions. Similarly, MMM has a positive bias in the Western Ghats of India for SI, which is mainly influenced by positive bias of P. The pattern of high SI over the Bay of Bengal is, however, not consistent among the GPCP, CMAP and MMM datasets, and follows the pattern of their precipitation maxima The MMM underestimates the observed P over the Ganges basin, while it overestimates P – and SI - over the eastern Himalaya. As of the coastal Mekong Basin, observations suggest high estimates of SI whereas MMM suggests the opposite, because of the lower estimates of RE. Similarly, for the eastern part of the Tibetan Plateau, simulated P is overestimated but SI estimates are similar to that of the observations because of the compensating negative RE bias.

Summarizing, models from MRI, IPSL, MPI, CMCC and MOHC along with seven models (NorESM1-M, INMCM4, FGOALS-g2, CNRM-CM5, CSIRO-Mk3-6-0, CanESM2 and bcc-csm1-1) substantially underestimate SI over most of the study domain, mainly as a result of underestimation of P. Five models (BNU-ESM, CESM-CAM5, EC-EARTH, MIROC-ESM-CHEM and MIROC-ESM), however, moderately underestimate SI over northern India (Ganges Basin), while only MIROC5 has a substantial positive bias. Only models from the NOAA-GFDL along with two models (CESM1-BGC and CCSM4) from NCAR suggest values of SI similar to observations. For most of the individual models, positive biases in RE compensate underestimation of P and SI, and vice versa. Instead, CanESM2 simulates low values of SI, associated to negative biases in P and in RE, the latter due to less concentrated MPR over most of the land region within the study domain.

**3.3 Future change in Monsoon onset, duration and RFA slope**

We present in Table 2 results of our investigation regarding the future projections from all models considered in the present-day analysis. However, in the text, we restrict our discussion to only the subset of models that provide a satisfactory representation of the present-day climatology and project statistically significant changes relative to the historical period for the select monsoonal metrics.

We note that most of the models project statistically insignificant changes in the monsoon onset, in active monsoon duration and in the RFA slope for the Indus basin. When looking at those models that project significant change for at least one of the select metrics, four models (BNU-ESM, CMCC-CESM, MIROC-ESM-CHEM and MPI-ESM-LR) foresee timing of the monsoon onset delayed by 1-3 pentads for the latter half of the 21st century, while three models (CanESM2, FGOALS-g2 and HadGEM2-ES) suggest an earlier onset by 1-2 pentads. The majority of models (BNU-ESM, CMCC-CM, GFDL-CM3, IPSL-CMA-MR, NorESM1-M and models from MOHC and MIROC) project a decrease of 2-6 pentads in the monsoon's duration, while only two models (CanESM2 and GFDL-ESM2M) suggest an increased duration of the monsoon by 4 pentads. Looking at the RFA slope, we find that the majority of models foresee a significant positive increase under future climate conditions, with changes ranging between 10 and 40%.

Similar to the Indus basin, most of the models project statistically insignificant changes for the Ganges basin. However, in contrast to the Indus basin, models that project significant changes agree well on a delayed onset by 2-4 pentads while only CanESM2 projects an early onset by 2 pentads. Almost half of the models project a significant shortening of the monsoon duration by 3-7 pentads (Table 2) while only two models (MIROC5 and CMCC-CESM) foresee the opposite, suggesting that it will be extended by 2-3 pentads, respectively. Similarly, nine models project a significant increase in RFA slope ranging between 10 and 30% while two models (MIROC5 and CMCC-CESM) foresee a reduction of about 10%.

For the Brahmaputra basin, models' agreement is weaker regarding change in the timings of the monsoon onset, as five models (BNU-ESM, CESM1-BGC, IPSL-CMA-MR, NorESM1 and MPI-ESM-LR) project a delayed onset by 1-4 pentads while four models (CSIRO-Mk3-6-0, CESM1-CAM5, INMCM4 and MIROC5) project the onset early by 1-2 pentads. However, almost half of the studied models agree well on a significant shortening of the monsoon duration by 2-7 pentads while only two models (CSIRO-Mk3-6-0 and MIROC5) project its extension by 3 pentads. Changes in the RFA slope are projected to be modest: one-third of models foresee a significant increase in the RFA slope by 4-15%, whereas only two models (CSIRO-Mk3-6-0 and MIROC5) project its decrease by 9%.

Looking at the Mekong basin, one finds that models foresee a weaker impact of climate change on the properties of the monsoon. Twelve models (BNU-ESM, CanESM2, CNRM-CM5, MPI-ESM-LR, and models from NCAR, NSF-DOE-NCAR, CMCC and only ESM from NOAA) project delay in the monsoon onset by 1-2 pentads. Five models (CESM1-BGC, CMCC-CMS, CMCC-CM, CSIRO-Mk3-6-0 and MPI-ESM-LR) project that the monsoon duration will be shortened by 2-3 pentads, while two models (MIROC-ESM-CHEM and MIROC5) project its extension by 1-2 pentads. Six models (CESM1-BGC, CMCC-CMS, CSIRO-Mk3-6-0 and CNRM-CM5, IPSL-CMA-MR and MPI-ESM-LR) project rise in the RFA slope by less than 10% while three models (ESM from MIROC and NorESM1) project the opposite suggesting its drop by less than 5%. The rest of the models either project no or insignificant change for the chosen metrics for the basin. The models generally tend to project a delayed monsoon onset, shortened duration of the monsoon and rise in the RFA slope.

We found that MMM shows a little skill for the chosen metrics (3 out of 12) as compared to some individual (CSIRO-Mk3-6-0, MIROC5 and GFDL-ESM2G) or group of models (MPI models). Because, of this, and given the discrepancies found across models, we have a little confidence on the actual reliability of the future change suggested by MMM. The MMM projects a monsoon onset delayed by 1-2 pentads for the Mekong and Ganges basins, respectively while the onset will not change for the Indus and Brahmaputra basins. The MMM projects that the monsoon duration will get shortened by one pentad for the Mekong and Indus basins, by two pentads for the Brahmaputra basin and by four pentads for the Ganges basin. The MMM also projects rise in the RFA slope of 1-10% for all basins by the end of present century.

### 3.4 Future changes in mean precipitation, relative entropy and seasonality index

In Table 5 we present future changes in P, RE and SI indicators under an extreme emission scenario of RCP8.5 for the future period (20161-2100) relative to historical period (1961-2000) for all study basins. Additionally, in Table 6 we interpret the results of change in RE in terms of number of wet and dry days for all study basins. Figure 10 presents the spatial scale projected changes in precipitation, number of dry days and seasonality or intermittency of the considered precipitation regimes by the end of century as suggested by MMM. We note a less intermittent future WPR over the Indus and Ganges basins, associated with a projected decrease in P and increase in number of dry days within WPR up to a fortnight as suggested by most of the models (Table 6). Over the Brahmaputra and Mekong basins, the projected decrease in the erratic behavior of WPR is associated with an increase in P but decrease in number of wet days (increased RE) up to three weeks as foreseen by majority of models. The seasonality of MPR and annual precipitation regime will increase over all study basins, which are associated with general projected increase in both P and RE.

The MPI, IPSL and BNU models along with five models (EC-EARTH, GFDL-CM3, INMCM4, CESM1-BGC and CCSM4) project a slight decrease in westerly P fairly regularly across most of the study domain but suggest substantial decrease in number of wet days (increase of RE) within WPR. Models also project a slight decrease in the erratic behavior of WPR over the region, with exceptions of five models (INMCM4, CMCC-CM and GFDL-CM3 and models from MPI), which mainly project an increase. The MMM suggests a negligible increase in P and SI, though large changes are expected in RE, such as, a substantial increase in the number of dry days over the northern Pakistan (HKH ranges, upper Indus Basin), southern India, eastern China, Myanmar region and lower Mekong Basin and decrease over the southern Pakistan, northern India and parts of the eastern Tibetan Plateau (Fig. 10). An increase in the number of dry days for WPR over the HKH area might be of great relevance in terms of changes in the cryosphere, and subsequently for the timely meltwater availability downstream. The WPR is projected to be less intermittent slightly along the eastern Indian coast and over the regions where number of dry days are increased (increased RE).

There is no obvious relationship between projected changes in the duration of the monsoon and in the total precipitation. Four models (CanESM2, CMCC-CESM and ESM from MPI and NOAA-GFDL) mainly project that monsoonal P will decrease while its duration will get shortened (increase in number of dry days) over the Indian sub-continent. A slight increase in P over the Mekong region is instead projected by such models, except by the CanESM2. Only models from BCC, MOHC and FGOALS-g2 project substantial decrease in P over the Mekong region. Models from NCAR, CMCC-CM, CMCC-CMS and EC-EARTH, project an increase in P and shorter duration (increase in number of dry days) of the monsoonal regime, implying more concentrated P over most of the study region. The IPSL, BNU and BCC models and two models (FGOALS-g2 and MIROC5) suggest substantially extended wet season over Pakistan and India while MOHC models project the same only over India. As a result, almost all models foresee a slight increase of SI over the study region, with BCC model as an outlier. The MMM suggests that P will substantially increase over the Brahmaputra basin, Himalayan range and adjacent parts of the Tibetan Plateau, including the Western Ghats. The study domain east of the Ganges basin shows relatively higher increase in RE and SI that is mainly associated with changes in P. The SI over the southern India, particularly over the Western Ghats will be increased. The RE will decrease over the northwest India (western Ganges Basin) and mid of Pakistan (lower Indus Basin) though there is negligible change in SI.

On an annual time scale, a prolonged duration of precipitation regime and negligible change in the SI are projected over the lower Indus Basin and western Ganges Basin. However, for the rest of the domain SI will increase with an increase in RE, which together with an increased P implies a highly concentrated precipitation regime in future. While interpreting the estimates of RE in terms of number of dry/wet days, we note an increase in number of dry days over most of the study domain up to a month, except for the lower Indus and western Ganges basins, exhibiting a slight increase in the number of wet days – a signal of change dominated by WPR. This has relevance in terms of management of water resources and regarding the production of electric energy from the hydropower stations. We note that the annual pattern of P, RE and SI are dominated by changes in the monsoonal regime, suggesting similar patterns, though the magnitude of change is somewhat higher.

Interesting conclusions can be drawn regarding changes in the extent of the monsoon. The MMM suggests a westward extension over the northwestern India and Pakistan and northward over China under the extreme warming scenario of RCP8.5 (Fig. 11). Most of the individual models suggest only slight but similar changes in the extent of the monsoon. Nevertheless, MMM exhibits relatively better performance on a spatial scale than most of the models analyzed.

## 4 Summary, Discussions, and Conclusions

Highly variable water supply drives the socio-economic wellbeing in South and Southeast Asia and guarantees growth of the agrarian economies. However, the region is exposed to various climatic extremes and adverse impacts of climate change, which are drastically exacerbating altogether the existing pattern of socio-economic vulnerability in the region. Much needed at present is the efficient management of resources and adequate policy devising to mitigate or at least better cope with the expected changes. This needs an adequate knowledge of prevailing and future state of possible changes in the climate system under the global warming scenario and assessment of the associated pattern of change in the hydrology at regional to local scale. Climate science and modelling communities have invested immensely in such pathways leading to the latest development of global climate models in the fifth phase (CMIP5). In this study we have reviewed common and consistent systematic biases from the CMIP5 models over the study region and their improved skills relative to their predecessors, as reported so far. Also, using state of the art technique, we have analyzed how well the latest generation climate models included in the fifth assessment report of IPCC (IPCC, 2013) are able to describe the seasonality of precipitation regimes associated with two large scale circulation modes: monsoon system and westerly disturbances over the major river basins of South and Southeast Asia. First, we have investigated the skill of the models for each basin in reproducing observed time-dependent properties of the seasonal cycle from extreme GHG emission scenario RCP8.5. In this regard, we have studied the timing of the monsoon onset, the monsoon duration and rate of the monsoonal precipitation concentration (our main monsoonal metrics). The diversified skill found from the individual models for each study basin, and appropriateness of the relative instead of the absolute but distinct thresholds for each individual basin further endorsed the effectiveness of our basin scale analysis. Secondly, we have applied time- and threshold-independent indicators of seasonality, such as RE and SI along with P, over the study basins and over the spatial extent of study domain and separately for the MPR, WPR and annual time scales. We have reported general and consistent biases of the CMIP5 models in terms of the chosen metrics, and present their projected future changes by the end of the century under an extreme global warming scenario (RCP8.5).

For the present climate, we note that the CMAP observations suggest similar timings of the monsoon onset, retreat and RFA slope as of the GPCP dataset for all basins, except for the Brahmaputra basin where it suggests little deviations such as the onset early by one pentad, a duration prolonged by two pentads and an RFA slope underestimated by 5%. Such consistency between the two dataset has been shown for various metrics and over different land regions (Sperber et al., 2013; Sperber and Annamalai, 2014; Pascale et al., 2014a & b; Prakash et al., 2014), despite of their differences in the magnitude of precipitation. Neither any single model nor the MMM performs best for the select monsoonal metrics against the considered observations. Models feature diverse skills in reproducing the seasonal cycle of precipitation over the study basins.

Similar to the CMIP3 models (Hasson et al., 2014a), we note that some CMIP5 models simulate unrealistic seasonal cycle of precipitation. These are the BNU-ESM model for the Indus basin, the IPSL-CMB-LR model for the Ganges and Brahmaputra basins and the EC-EARTH model for all the basins. Hence, such models perform worst in terms of timing of the monsoon onset, monsoon duration and the RFA slope. Here, a surprising performance of the EC-EARTH model that features a second highest resolution among the CMIP5 models analyzed, indicate that without an adequate representation of the requisite physical climatic processes, only high resolution cannot achieve the realistic simulation over the region. One-third of the studied models (bcc-csm1-1, IPSL-CMA-LR, IPSL-CMB-LR, CSIRO-Mk3-6-0, FGOALS-g2, INMCM4 and models from CMCC and MRI) featuring either a smoothed or a gentle growth in their fractional accumulated P (low RFA slopes) fail to adequately simulate the monsoonal precipitation regime over the Indus basins (Fig. 5). Three of these models (IPSL, INMCM and MRI models) consistently show such discrepancy from their earlier versions participating in the CMIP3 archive (Hasson et al., 2014a).

Almost half of the models suggest right timings of the monsoon onset for the Ganges and Brahmaputra basins and slightly delayed onset for the Indus basin, while there is a mixed behavior for the Mekong basin. An early onset suggested by some models (e.g. over the Mekong basin) may possibly be linked to their simulated marked land sea temperature contrasts (Annamalai et al., 2007). We note that realistic onset timings over the Ganges basin by most of the studied models is in contrast to the general behavior of a delayed onset over India from the CMIP3 and CMIP5 models (Sperber et al., 2013; Sperber and Annamalai, 2014). This may be attributed to a marked spatial heterogeneity of the monsoonal precipitation, for which typically a uniform threshold is applied over the whole domain, for the onset timings. Therefore, we encourage probing models' skill for the monsoon related metrics over small units (such as river basins) and through applying distinct thresholds appropriate for such units.

On the other hand, a delayed onset over the Indus basin together with an underestimation of monsoon P is mainly linked to a well reported systematic error of suppression of the monsoon far north over China and far west over Pakistan, which is common amid both CMIP3 and CMIP5 modeling efforts (Boos and Hurley, 2013; Sperber et al., 2013; Hasson et al., 2013 and 2014a). In these models, an overly smoothed orography west of the Tibetan Plateau (60°E–80°E longitudinal band), constrained by models' horizontal resolutions, allows intrusion of the mid-latitude dry air well into the monsoonal thermal regime (Chakraborty et al., 2002 and 2006). This bogus penetration of cold air weakens the upper-tropospheric thermal maximum, shifts it towards the southeast and suppresses the moist convection (Boos and Hurley, 2013). Hence, strength of the monsoon P and its low level jet subsides, preventing the monsoonal regime extending too far to its western and northwestern extremity, resulting over there a delayed onset and a negative P anomaly (Hoskins and Rodwell 1995; Chakraborty et al., 2002 and 2006).

Similar to the underrepresentation of the realistic Himalaya-Karakoram orography, absence of the irrigation schemes in models considerably contributes to the systematic monsoon bias over the region (Syed et al., 2009 and 2013) that result in an underestimation of P and a delayed onset. The irrigation is particularly relevant for the Indus and Ganges basins, where annually a large amount of water is diverted to and evaporated from the agricultural fields (Hasson et al., 2013). Syed et al. (2009) show that significant model biases of temperature and mean sea level pressure over parts of the Indus basin are sensitive to the water used for irrigation over there. They also show that representation of irrigation scheme in models can result in a relatively small land-sea thermal contrast over parts of the Indus basin during summer, reducing penetration of the westerlies from the Arabian Sea. This creates favorable conditions for the monsoonal currents originating from the Bay of Bengal to penetrate well and deeper into the west and northwest India and Pakistan. Similarly during the winter season, the CMIP3 and CMIP5 coupled models feature a cold sea surface temperature (SST) bias over the northern Arabian Sea that persists into spring and summer seasons (Levine and Turner 2012; Levine et al., 2013). As a consequence, the low-level monsoonal jet features smaller amount of moisture due to less evaporation over cooler north Arabian Sea that subsiding the monsoon convection and low-level convergence over land reduces the strength of the monsoonal flow and subsequently the P (Levine et al., 2013). Marathayil et al. (2013) describe that in the CMIP3 models, the cold SST bias over north Arabian Sea mainly results from the advection of cold/dry air by anomalously stronger north-easterlies and colder surface temperatures simulated by the models over Pakistan and northwest India. Recently, Sandeep and Ajayamohan (2014) have shown looking at CMIP5 models data that the equatorward bias in the subtropical Jetstream is responsible for such anomalous cooling of SST over the north Arabian Sea. These systematic biases in the models, either due to absence or underrepresentation of important features, makes the realistic simulation of climate over the Indus and Ganges basins extremely difficult.

Models generally feature a systematic bias of simulating extended monsoon duration relative to the observations for the Indus, Ganges and Brahmaputra basins. This is mainly due to fact that models fail to simulate the observed seasonality, so the RFA slope is generally underestimated. The retreat threshold is, therefore, achieved later than its observed timings, suggesting extended active monsoon duration for such basins. On the other hand, models tend to simulate shortened monsoon duration for the Mekong basin, which is due to higher precipitation concentration (overestimated RFA slope) against the observations. Interestingly, we note that models suggesting the RFA slope similar to that of the observations simulate mostly a realistic active duration of the monsoon. Relevant examples are five models (GFDL-ESM2G and GFDL-ESM2M, MPI-ESM-LR, MPI-ESM-MR and MIROC5) for the Ganges basin, two models (CSIRO-Mk3-6-0 and MPI-ESM-MR) for the Brahmaputra basin and five models (CCSM4, GFDL-CM3, MIROC-ESM-CHEM, MIROC-ESM, MIROC5 and NorESM1-M) for the Mekong basin (Table 2). Otherwise, the RFA slope is highly correlated negatively to the monsoon duration for any particular model. Hence, the higher the underestimation of an RFA slope, the larger the overestimation of active monsoon duration. In few cases, RFA slope is also associated with the suggested timing of the onset. For instance, the RFA slope is underestimated in a case when onset threshold is achieved earlier than the pentad when actually the rapid accumulation starts. This situation occurs when the models excessively simulate P during the pre-monsoonal period, suggesting wrong timing of the onset. Examples are the nine models (bcc-csm1-1, FGOALS-g2, IPSL-CMB-LR and models from MOHC and MRI) over the Ganges basin. Sperber and Annamalai (2014) explain that solo heavy rainfall events can result into such a bogus onset.

In summary, one-third of the analyzed models performed well for multiple select metrics (Table 2), despite of the differences in the magnitude of their simulated precipitation. Considering a total of 12 metrics (3 metrics x 4 basins), the MPI models perform well for 7 metrics, two models (MIROC5 and CSIRO-Mk3-6-0) perform well for 5 metrics, and six models (CCSM4, CESM1-CAM5, GFDL-ESM2G, IPSL-CMA-MR, MIROC-ESM and MIROC-ESM-CHEM) perform well for 4 metrics. Most of these models show their skill mainly for the Ganges basin, followed by the Brahmaputra, Mekong and Indus basins, respectively. Our results of better representation of seasonal cycle of monsoonal P for MIROC and CSIRO-Mk3-6-0 models (5 out of 12 metrics) are in agreement with Babar et al. (2014).

For the basin-integrated seasonality indicators, we found that most of the models overestimate (underestimate) the extent of concentration of P for MPR and number of dry days for WPR (RE) for the Indus and Ganges Basins while they underestimate (overestimate) them for the Brahmaputra and Mekong Basins. However, simulations provide a positive bias in SI of WPR in all basins, while positive (negative) biases for MPR are found for the Ganges and Brahmaputra (Indus and Mekong) basins, following the pattern of overestimation (underestimation) of precipitation. It is pertinent to mention here that seasonality of the MPR (SI) positively correlates with the RFA slope estimate, despite the fact that the RFA slope is calculated within the active monsoon duration while SI accounts for the whole wet season, including precipitation from the pre-onset and post-retreat monsoon season. However, since RFA slope estimates are sensitive to identification of right timings of the monsoon onset, as discussed earlier, its relation with SI should be carefully considered.

Here, we emphasize that overestimation of the westerly P over the study basins needs a careful interpretation, particularly for the Indus basin, where a large mountainous part of the basin in the north that is mainly affected by the westerly P, features a very sparse observational network and only few, valley-based, rainfall-only, stations are incorporated in the global gridded datasets. On the other hand, merged estimates from the satellite datasets as in case of GPCP/CMAP are largely affected by certain limitation of estimation of precipitation in such a high relief area (Palazzi et al., 2013). In view of increasing number of observatories in HKH region (Hasson et al., 2014b), we encourage validation of the simulated precipitation additionally against newly available station observations, in order to be more confident about the climate model performance in such high relief areas.

For the seasonality indicators over whole spatial domain (as Fig. 1), no single model was found satisfactory against observations. The MMM though does not outperform all individual models but somewhat provides a fair agreement with the observations with few exceptions (Table 4). For WPR, most of the models generally overestimate SI due to overestimation of P and relatively higher (lower) number of wet days over the high (low) land areas. MIROC ESMs followed by models from NSF-DOE-NCAR suggest highest P for WPR over the region. In contrast to the general underestimation of the monsoonal P over the Indian plains, models from NCAR, NSF-DOE-NCAR and NOAA-GFDL along with EC-EARTH suggest right magnitude, even though the latter model simulates an unrealistic seasonal cycle of precipitation. Overall, models feature large biases for P and for its spatio-temporal distribution. Models generally suggest large RMSE of P and RE for all considered time scales but a high spatial correlation (PC >= 0.7) for P, RE and SI against the observations. Most of the models suggest σ of P within the observational uncertainty for the monsoon duration and on annual time scale while they overestimate it for the WPR. For the monsoon season, such results are consistent with findings of Collins et al. (2013), who have considered a larger spectrum of the observational uncertainty.

For the future climate, we note that two of IPSL low resolution models suggest statistically significant changes in the select metrics (Table 2). However, we have a little confidence on such changes since these models have poor performances over the Ganges and Brahmaputra basins during the historical period. Similar is the case with EC-EARTH for all basins and BNU-ESM for the Indus basin. We note that models projecting an early (delayed) onset in future also project extended (shortened) monsoon duration. Similarly, if the monsoon duration is projected to be extended (shortened), the RFA slope is projected to decrease (increase). Most of the models suggesting significant changes agree well on a delayed onset by 2-4 pentads for the Ganges and by 1-2 pentads for the Mekong basin but there is a mixed response of delayed or early onset by 1-4 pentads for the Indus and Brahmaputra basins. The projected delayed onset might be influenced by a profound effect of underrepresentation of topography and/or absence of irrigation water as suggested by studies (Syed et al., 2009; Chakraborty, 2006; Levine and Turner 2012; Marathayil et al., 2013; Levine et al., 2013; Boos and Hurley, 2013). The models also agree well on the shortening of the monsoon duration by 2-7 pentads and on the rise of the RFA slope by 4-40% for all study basins – projected rise in the RFA slope shows a positive gradient from the Mekong to the Indus basin. The shortening of the monsoon duration is in contrast to Kiripalani et al. (2007) who have suggested its possible extension, but is consistent with the recent observations (Ramesh and Goswami, 2007). It is pertinent to mention here that signal of a general shortening of the active monsoon duration depends upon the choice of relative thresholds - particularly one for the monsoon retreat - which are commonly applied to the present and future climates. Given that the future climates are attributed to increase in precipitation - as apparent from the results for most of the study region - the fractional threshold for the monsoon retreat tends to be reached earlier as compared to the present climate. Therefore, suggesting shortening of the active monsoon duration in future here is merely relative to the present climate and actually suggests the more concentrated monsoonal precipitation regime. Also, in view of the poor performance from the models in reproducing the observed length of the active monsoon duration during the present climate, we emphasize on the change in concentration of precipitation, instead of change in the duration.

We note that most of the suggested changes in the timings of onset are of few pentads, which signifies the effectiveness and use of a fine-grained dataset for such analysis. This is why Hasson et al. (2014a) have shown almost no change for the CMIP3 models on a monthly time scale in the onset timings of the monsoon by the end of century. The suggested changes, though small, may have stark impact on the local hydrology and agricultural regions, particularly over the semiarid and arid plains, and as a whole, on the country scale economic conditions.

We have found a less intermittent WPR and higher seasonality of MPR and annual precipitation regimes over all study basins in future, which are generally associated with a projected increase in both P and extent of its concentration for MPR and number of dry days within WPR (increase in RE), except for the Indus and Ganges basins for which P will decrease for WPR. Our findings of a less intermittent future WPR for the Indus basin, associated with an increase in the number of dry days and decrease in P, is consistent with the observed drying of the spring season over the HKH region within the Upper Indus Basin (Hasson et al., 2015 – in preparation) and partly with a more frequent occurrence of westerly disturbances over the Karakoram (Ridley et al., 2013). Such changes are mainly responsible for the ongoing reduction in the ephemeral snowpack extent therein (Hasson et al., 2014), and subsequent observed change in the seasonal water availability. Moreover, future decrease in P under WPR may result in a negative budget for the UIB cryosphere, posing a serious threat for much needed melt-water by the arid region downstream. A decrease in P under WPR is linked to a poleward shift of the westerly storm track as reported by various studies (Bengtsson et al., 2006; Fu, 2006; Fu and Lin 2011). Furthermore, projected increase in the monsoonal P and extent of its concentration indicate intensification of MPR under future warming, which may be associated with the extreme hydro-meteorological conditions, already projected (Hirabayashi et al. 2013) and evident from the observations (e.g. a series of 2010, 2012 and 2014 monsoonal floods in Pakistan). It is to mention that monsoon breaks are a crucial phenomenon, which severely affects the rain fed agricultural areas in the region. The statistics of monsoon breaks can largely affect the RE estimates for the MPR. Thus, performance of the models in better simulating the MPR and the reliable projected changes in RE largely depend upon the simulated statistics of the monsoonal breaks, which we have not explicitly interpreted here. We open a further discussion on how the projected changes in the MPR presented here can be cross checked by separately taking into account statistics of the monsoonal breaks and its effect on the measure of RE and its explicit interpretation, while studying the changes in the monsoon seasonality or length of the wet season.

In contrast to the observed decrease in the spatial extent of the south Asian summer monsoon (Ramesh and Goswami, 2007), we have found from MMM its extension, which is northward over China and westward over northwest India and Pakistan. Such results are consistent with Lee and Wang, (2014), who also suggest westward shifts in the monsoonal domain and with Kitoh et al. (2013), who also show small changes in the monsoonal domain over the region. This implies that the border areas will experience critical changes in their precipitation regime in future (Seth et al. 2013; Hasson et al. 2014a). It is pertinent to mention here that due to the limited skill of the models in reproducing the monsoonal regime over the northwest India and Pakistan such MMM based projection of changes in the monsoonal domain westward owes a little confidence. It is therefore necessary to see such a change from set of 'reasonable' individual models that feature minimum biases over the northwest India and Pakistan such as CCSM4, CESM1-BGC, EC-EARTH and GFDL-CM3 (see supplement Fig. 2).

Various discrepancies in representation of the seasonal cycle of precipitation in the CMIP5 models are mainly associated with the representation of the south Asian summer monsoon. Such discrepancies generally attribute to issues with large scale atmospheric circulations, underrepresentation of real orography, and simplest form of land-atmospheric-ocean processes and their interaction. We emphasize here that inclusion of the irrigation water and appropriate representation of the orography are vital for realistically reproducing the summer monsoonal precipitation regime and for its reliable future changes over the region. The dynamical downscaling using Regional Climate Models (RCMs) - though computationally expensive and largely depends upon skill of the global model forcing - can be helpful in improving simulation of the monsoonal regime by incorporating the local-scale geo-physical characteristics and detailed land-use/-cover dynamics through achieving high resolutions (Hasson et al., 2013; Palazzi et al., 2013). In view of the diverse skillset found for the CMIP5 models, we suggest that use of their output in further impact assessment models and for policy making in the region should base on only a set of 'reasonable' models in that particular aspect (Saha et al., 2014), and preferably be supported by the fine-scaled dynamical downscaling efforts, such as Coordinated Regional Climate Downscaling Experiment (CORDEX) South Asia. We also conclude that Indus and Ganges basins (Pakistan and north-/west India region) are very critical in nature and difficult for the present-day climate models in order to reproduce their climate. This has subsequent implications for driving the impact assessment models for assessing climate change impacts on various socio-economic development sectors for these basins. In such regards, the state-of-the-art coupled models need to be improved enormously and meaningfully, particularly for the representation of region-specific geo-physical characteristics and their interaction with the physical processes that are presently absent completely or represented inadequately, so far.

*Acknowledgements*: The authors acknowledge the World Climate Research Programme's Working Group on Coupled Modelling, which is responsible for CMIP, and we thank the climate modeling groups (listed in Table 01 of this paper) for producing and making available their model output. For CMIP the U.S. Department of Energy's Program for Climate Model Diagnosis and Intercomparison provides coordinating support and led development of software infrastructure in partnership with the Global Organization for Earth System Science Portals. SH and JB acknowledge the support of BMBF, Germany's Bundle Project CLASH/Climate variability and landscape dynamics in southeast Tibet and the eastern Himalaya during the Late Holocene reconstructed from tree rings, soils and climate modelling. VL and SP acknowledge the support of the FP7/ERC Starting Investigator grant NAMASTE/Thermodynamics of the Climate System (Grant No. 257106). The authors also acknowledge the support from CliSAP/Cluster of excellence in the Integrated Climate System Analysis and Prediction.

**Table 1. List of CMIP5 models, their modelling groups and resolutions.**

| S. No. | Modeling Group | Model Name | Atm. Resolution (lonxlat) | Model Level |
|---|---|---|---|---|
| 1 | Beijing Climate Center, China Meteorological Administration(BCC) | BCC-CSM1-1 | 2.8 x 2.8 | 26 |
| 2 | College of Global Change and Earth System Science, Beijing Normal University (GCESS) | BNU-ESM | 2.8 x 2.8 | 32 |
| 3 | Canadian Centre for Climate Modelling and Analysis (CCCMA) | CanESM2 | 2.8 x 2.8 | 35 |
| 4 | NCAR Community Climate System Model, (CCSM) | CCSM4 | 1.25 x 0.94 | 27 |
| 5 | Community Earth System Model Contributors | CESM1-BGC | 1.25 x 0.94 | 27 |
| 6 | (NSF-DOE-NCAR) | CESM1-CAM5 | 1.25 x 0.94 | 27 |
| 7 | Centro Euro-Mediterraneo per I Cambiamenti Climatici (CMCC) | CMCC-CESM | 3.75 x 3.75 | 39 |
| 8 | | CMCC-CMS | 1.875 x 1.875 | 95 |
| 9 | | CMCC-CM | 0.75 x 0.75 | 31 |
| 10 | Centre National de Recherches Météorologiques Centre Européen de Recherche et Formation Avancée en Calcul Scientifique | CNRM-CM5 | 1.4 x 1.4 | 31 |
| 11 | Commonwealth Scientific and Industrial Research Organization in collaboration with QCCCE (CSIRO-QCCCCE) | CSIRO-Mk3-6-0 | 1.875 x 1.875 | 18 |
| 12 | EC-EARTH consortium (EC-EARTH) | EC-EARTH | 1.125 x1.125 | 62 |
| 13 | LASG, Institute of Atmospheric Physics, Chinese Academy of Sciences and CESS, Tsinghua University (LASG-CESS) | FGOALS-g2 | 2.8125 x 2.8125 | 26 |
| 14 | NOAA Geophysical Fluid Dynamics Laboratory (NOAA-GFDL) | GFDL-CM3 | 2.5 x 2.0 | 48 |
| 15 | | GFDL-ESM2G | 2.5 x 2.0 | 24 |
| 16 | | GFDL-ESM2M | 2.5 x 2.0 | 24 |
| 17 | Met Office Hadley Centre (MOHC) | HadGEM2-CC | 1.875 x 1.24 | 60 |
| 18 | | HadGEM2-ES | 1.875 x1.24 | 38 |
| 19 | Institute for Numerical Mathematics (INM) | INMCM4 | 2 x1.5 | 21 |
| 20 | Institut Pierre-Simon Laplace (IPSL) | IPSL-CM5A-LR | 3.75 x1.89 | 39 |
| 21 | | IPSL-CM5A-MR | 2.5 x1.25 | 39 |
| 22 | | IPSL-CM5B-LR | 3.75 x1.9 | 39 |
| 23 | Japan Agency for Marine-Earth Science and Technology, Atmosphere and Ocean Research Institute (The University of Tokyo), and National Institute for Environmental Studies (MIROC) | MIROC-ESM-CHEM | 2.8 x2.8 | 80 |
| 24 | | MIROC-ESM | 2.81 x 2.81 | 80 |
| 25 | | MIROC5 | 1.4 x 1.4 | 40 |
| 26 | Max-Planck-Institut für Meteorologie (MPI-M) | MPI-ESM-LR | 1.875 x1.875 | 47 |
| 27 | | MPI-ESM-MR | 1.875 x1.875 | 95 |
| 28 | Meteorological Research Institute (MRI) | MRI-CGCM3 | 1.125x1.125 | 48 |
| 29 | | MRI-ESM1 | 1.125 x 1.125 | 48 |
| 30 | Norwegian Climate Centre (NCC) | NorESM1-M | 2.5 x 1.9 | 26 |

**Table 2.** Offset in the timings of monsoon onset (O) and its duration (D) in pentads and % slope of the rapid fractional accumulation (S) for historical period (1961-2000) relative to the GPCP data and change for the future period (2061-2100) relative to the historical period. Note: Negative values imply decrease or early timings while positive values suggest the opposite. The bold values for the historical period (1961-2000) suggest statistically insignificant offset from observations while for the RCP8.5 bold figures suggest statistically significant changes.

| | HISTORICAL (1961-2000) | | | | | | | | | | | | RCP8.5 (2061-2100) | | | | | | | | | | | |
|---|---|---|---|---|---|---|---|---|---|---|---|---|---|---|---|---|---|---|---|---|---|---|---|---|
| Data | INDUS | | | GANGES | | | BRAHMA | | | MEKONG | | | INDUS | | | GANGES | | | BRAHMA | | | MEKONG | | |
| | O | D | S | O | D | S | O | D | S | O | D | S | O | D | S | O | D | S | O | D | S | O | D | S |
| GPCP | 37 | 16 | 0.05 | 32 | 25 | 0.04 | 29 | 28 | 0.03 | 26 | 36 | 0.025 | | | | | | | | | | | | |
| CMAP | **0** | **2** | -8 | **1** | **-1** | 5 | **-1** | **2** | -5 | **0** | **-1** | 3 | | | | | | | | | | | | |
| MMM | **1** | 10 | -45 | -4 | 10 | -24 | -2 | 7 | -16 | **-1** | **0** | 5 | 0 | -1 | 8 | 2 | **-4** | **10** | 0 | **-2** | 5 | 1 | -1 | 1 |
| bcc-csm1-1 | 8 | 9 | -46 | -8 | 21 | -47 | **-1** | 11 | -21 | 1 | -5 | 13 | -1 | 1 | -4 | 2 | -3 | 4 | 0 | -1 | 1 | 0 | 0 | -2 |
| BNU-ESM | 3 | 10 | -46 | **2** | 4 | -13 | 2 | 3 | -6 | **0** | **-1** | 7 | **2** | **-3** | **21** | **3** | **-4** | **7** | **2** | **-2** | 4 | **2** | -1 | 3 |
| CanESM2 | **2** | 8 | -34 | **1** | 6 | -20 | 1 | 3 | -8 | -2 | 4 | -7 | **-2** | **4** | **-21** | -2 | 1 | 1 | 0 | **-3** | **15** | 1 | -2 | 4 |
| CCSM4 | **1** | 6 | -28 | **1** | 6 | -21 | -1 | 6 | -21 | -1 | **2** | **0** | 0 | -1 | -1 | **2** | **-3** | **8** | 0 | **-2** | **8** | **2** | 0 | 0 |
| CESM1-BGC | **1** | 6 | -30 | **1** | 5 | -16 | -1 | 6 | -21 | -2 | **2** | **-1** | 0 | -1 | 1 | **2** | **-3** | **11** | **1** | **-2** | **7** | **2** | **-2** | 4 |
| CESM1-CAM5 | **0** | 5 | -22 | **1** | **1** | -8 | -3 | 7 | -22 | -2 | 2 | **-1** | 0 | 1 | -6 | 0 | 0 | 2 | **-1** | 0 | 4 | **2** | -1 | 2 |
| CMCC-CESM | 2 | 10 | -50 | -4 | 5 | -12 | **-2** | 4 | -9 | **0** | **-1** | 6 | **3** | -1 | 7 | -2 | **3** | **-10** | 1 | **-3** | 5 | **2** | -1 | 2 |
| CMCC-CMS | 3 | 12 | -53 | **-3** | 10 | -25 | -1 | 5 | -12 | **0** | -2 | 10 | 0 | -2 | **17** | **3** | **-6** | **18** | -1 | -2 | 5 | **2** | **-3** | 6 |
| CMCC-CM | 2 | 13 | -53 | **0** | 8 | -23 | **0** | 3 | -5 | 1 | -4 | 14 | **1** | **-4** | **21** | 1 | **-5** | **26** | 1 | **-3** | **9** | **1** | **-3** | 4 |
| CNRM-CM5 | **1** | 10 | -42 | **0** | 5 | -16 | -3 | 7 | -17 | -3 | 5 | -8 | 0 | 0 | 2 | 0 | -1 | 3 | 0 | -1 | 1 | 1 | -1 | 4 |
| CSIRO-Mk3-6-0 | **0** | 12 | -55 | **-1** | 5 | -13 | 1 | 1 | 2 | 2 | -7 | 15 | 1 | 0 | 2 | 4 | -2 | 6 | -2 | **3** | **-9** | 1 | -2 | 7 |
| EC-EARTH* | -2 | 15 | -54 | -19 | 29 | -60 | -11 | 19 | -44 | -20 | 27 | -52 | 0 | 0 | -1 | -1 | 1 | -5 | -2 | 3 | -9 | 0 | 0 | -3 |
| FGOALS-g2 | 2 | 10 | -47 | -9 | 18 | -39 | -3 | 8 | -19 | **0** | **1** | 2 | -2 | -1 | **11** | **7** | **-10** | **23** | **2** | **-3** | **7** | **-4** | **8** | **-18** |
| GFDL-CM3 | 3 | 7 | -34 | 2 | 6 | -19 | -2 | 11 | -25 | 2 | **-1** | 5 | -1 | -2 | **12** | 0 | **-4** | **16** | -1 | -2 | 1 | 0 | 0 | 3 |
| GFDL-ESM2G | 3 | 6 | -30 | 4 | **-2** | **0** | **0** | 5 | -14 | 1 | -3 | 9 | 0 | 0 | -2 | 0 | 1 | -3 | 0 | -1 | 2 | **1** | 1 | -2 |
| GFDL-ESM2M | 3 | 5 | -27 | 3 | **0** | -4 | -1 | 6 | -17 | 2 | -4 | 15 | -1 | 0 | -4 | 1 | **-2** | **7** | 1 | -1 | 2 | | | |
| HadGEM2-CC | -2 | 11 | -47 | -9 | 15 | -30 | -6 | 8 | -18 | 2 | -4 | 16 | 0 | **-5** | **38** | 4 | **-9** | **23** | 0 | **-4** | **8** | 1 | 0 | -1 |
| HadGEM2-ES | -1 | 12 | -49 | -9 | 14 | -29 | -6 | 8 | -17 | 3 | -5 | 19 | -1 | **-5** | **34** | **5** | **-10** | **27** | 1 | **-4** | **9** | 0 | 1 | -5 |
| inmcm4 | 2 | 11 | -49 | **-1** | 11 | -29 | **0** | 6 | -16 | -4 | 8 | -14 | 0 | -1 | 5 | **2** | **-5** | **15** | -1 | -1 | 4 | -1 | 0 | 0 |
| IPSL-CM5A-LR* | 4 | 12 | -55 | -11 | 21 | -42 | -5 | 12 | -24 | -2 | **1** | **0** | -2 | 1 | 8 | **13** | **-16** | **60** | **9** | **-12** | **43** | **3** | **-3** | **9** |
| IPSL-CM5A-MR | **1** | 14 | -60 | **-3** | 11 | -26 | -3 | 8 | -18 | -2 | **0** | 1 | 0 | **-3** | **38** | 4 | **-7** | **29** | **4** | **-7** | **20** | **3** | **-3** | 6 |
| IPSL-CM5B-LR* | -2 | 19 | -67 | -23 | 36 | -68 | -11 | 21 | -41 | 4 | -6 | 31 | 0 | -1 | 5 | **3** | **-4** | **31** | **3** | **-6** | **13** | **2** | -1 | 2 |
| MIROC-ESM-CHEM | -1 | 9 | -41 | **-1** | 9 | -30 | **0** | 7 | -23 | 2 | **1** | **0** | 1 | -2 | **16** | **2** | **-4** | **13** | 0 | -1 | 4 | 0 | **1** | -3 |
| MIROC-ESM | -2 | 10 | -43 | **-2** | 8 | -28 | **0** | 6 | -22 | 2 | **1** | **0** | 1 | **-4** | **27** | **2** | **-4** | **14** | -1 | 0 | 3 | 0 | 0 | -2 |
| MIROC5 | **0** | 12 | -49 | **0** | **-1** | 2 | -2 | 3 | -12 | -3 | **1** | -5 | 0 | **-6** | **33** | 1 | 2 | -9 | -2 | **3** | **-9** | -1 | **2** | -1 |
| MPI-ESM-LR | **1** | 10 | -45 | **0** | **2** | -7 | **0** | **2** | -6 | **-1** | -2 | 7 | 0 | -2 | 9 | **4** | **-3** | **6** | **2** | **-2** | **6** | 1 | **-3** | 7 |
| MPI-ESM-MR | 2 | 9 | -42 | **0** | **2** | -5 | 1 | **2** | -4 | 1 | -4 | 13 | **2** | 0 | 2 | **3** | -1 | 3 | 0 | 0 | 1 | 1 | -1 | 1 |
| MRI-CGCM3* | **0** | 14 | -55 | -10 | 19 | -42 | -2 | 5 | -9 | 3 | -5 | 24 | -1 | 1 | -1 | -1 | 0 | 4 | **-2** | 0 | -2 | **1** | -1 | 1 |
| MRI-ESM1* | **-1** | 14 | -55 | -11 | 21 | -44 | -2 | 4 | -8 | 2 | -5 | 26 | 0 | 1 | 1 | **3** | **-2** | **8** | 1 | -4 | 4 | **2** | 0 | 0 |
| NorESM1-M | 2 | 7 | -33 | **1** | 6 | -18 | 2 | 2 | -8 | **0** | **0** | 5 | 1 | **-3** | **12** | **4** | **-5** | **10** | **1** | **-2** | **4** | 0 | 1 | -5 |

**Table 3. Estimates for the basin integrated seasonality indicators, Precipitation (P), Relative Entropy (RE) and Seasonality Index (SI) for the historical period (1961-2000).**

| Basins→ | Indus | | | | | | Ganges | | | | | | Brahmaputra | | | | | | Mekong | | | | | |
|---|---|---|---|---|---|---|---|---|---|---|---|---|---|---|---|---|---|---|---|---|---|---|---|---|
| | Westerly | | | Monsoon | | | Westerly | | | Monsoon | | | Westerly | | | Monsoon | | | Westerly | | | Monsoon | | |
| Model and Obs.↓ | P | RE | SI | P | RE | SI | P | RE | SI | P | RE | SI | P | RE | SI | P | RE | SI | P | RE | SI | P | RE | SI |
| GPCP | 117 | 1.6 | 0.05 | 296 | 1.3 | 0.12 | 93 | 2.1 | 0.06 | 874 | 0.9 | 0.24 | 140 | 1.4 | 0.07 | 876 | 0.6 | 0.15 | 256 | 1.4 | 0.12 | 1292 | 0.5 | 0.21 |
| CMAP | 117 | 1.8 | 0.06 | 244 | 1.4 | 0.10 | 81 | 2.2 | 0.05 | 735 | 0.9 | 0.21 | 144 | 1.4 | 0.06 | 683 | 0.5 | 0.11 | 216 | 1.5 | 0.10 | 1118 | 0.5 | 0.18 |
| bcc-csm1-1 | 262 | 1.6 | 0.14 | 151 | 2.0 | 0.03 | 253 | 1.9 | 0.23 | 592 | 1.8 | 0.10 | 403 | 0.9 | 0.21 | 1542 | 0.7 | 0.13 | 237 | 1.3 | 0.19 | 1994 | 0.8 | 0.20 |
| BNU-ESM | 199 | 1.8 | 0.14 | 265 | 1.4 | 0.11 | 177 | 2.0 | 0.19 | 961 | 0.9 | 0.27 | 298 | 1.0 | 0.17 | 1289 | 0.5 | 0.22 | 285 | 1.0 | 0.17 | 1291 | 0.3 | 0.13 |
| CanESM2 | 112 | 2.0 | 0.09 | 177 | 1.3 | 0.07 | 125 | 2.2 | 0.16 | 706 | 0.8 | 0.17 | 221 | 1.2 | 0.14 | 1217 | 0.5 | 0.21 | 343 | 0.9 | 0.17 | 1124 | 0.3 | 0.11 |
| CCSM4 | 203 | 1.9 | 0.15 | 514 | 1.2 | 0.19 | 244 | 2.1 | 0.29 | 1265 | 0.9 | 0.34 | 497 | 1.1 | 0.30 | 1919 | 0.5 | 0.29 | 358 | 0.9 | 0.19 | 1292 | 0.4 | 0.18 |
| CESM1-BGC | 201 | 1.9 | 0.15 | 488 | 1.2 | 0.18 | 228 | 2.1 | 0.27 | 1261 | 0.9 | 0.35 | 514 | 1.1 | 0.30 | 1884 | 0.5 | 0.28 | 345 | 1.0 | 0.20 | 1266 | 0.4 | 0.17 |
| CESM1-CAM5 | 152 | 1.9 | 0.12 | 447 | 1.1 | 0.14 | 143 | 2.4 | 0.19 | 1340 | 0.7 | 0.30 | 764 | 1.1 | 0.41 | 2338 | 0.4 | 0.28 | 351 | 1.1 | 0.21 | 1139 | 0.4 | 0.14 |
| CMCC-CESM | 136 | 2.2 | 0.11 | 104 | 2.2 | 0.05 | 130 | 2.1 | 0.14 | 652 | 0.8 | 0.13 | 206 | 1.4 | 0.17 | 916 | 0.5 | 0.13 | 248 | 1.4 | 0.20 | 1297 | 0.3 | 0.12 |
| CMCC-CM | 249 | 2.4 | 0.18 | 192 | 2.4 | 0.09 | 180 | 2.8 | 0.20 | 714 | 1.4 | 0.27 | 479 | 1.3 | 0.29 | 1820 | 0.7 | 0.36 | 150 | 2.1 | 0.18 | 1035 | 0.6 | 0.21 |
| CMCC-CMS | 288 | 2.1 | 0.19 | 194 | 2.3 | 0.09 | 216 | 2.6 | 0.24 | 887 | 1.1 | 0.26 | 489 | 1.1 | 0.28 | 1712 | 0.5 | 0.28 | 197 | 1.8 | 0.20 | 1117 | 0.6 | 0.20 |
| CNRM-CM5 | 246 | 1.5 | 0.16 | 334 | 1.1 | 0.11 | 176 | 2.1 | 0.18 | 922 | 0.8 | 0.21 | 472 | 1.0 | 0.28 | 1587 | 0.5 | 0.23 | 380 | 0.9 | 0.21 | 1163 | 0.4 | 0.15 |
| CSIRO-Mk3-6-0 | 102 | 1.9 | 0.08 | 79 | 2.1 | 0.03 | 97 | 2.3 | 0.11 | 487 | 1.9 | 0.18 | 241 | 1.3 | 0.15 | 1253 | 0.5 | 0.19 | 148 | 1.6 | 0.13 | 1464 | 0.6 | 0.27 |
| EC-EARTH | 170 | 1.6 | 0.13 | 598 | 0.7 | 0.13 | 147 | 2.1 | 0.17 | 1048 | 0.4 | 0.15 | 315 | 1.1 | 0.18 | 1174 | 0.3 | 0.10 | 376 | 1.0 | 0.20 | 928 | 0.3 | 0.10 |
| FGOALS-g2 | 268 | 1.0 | 0.12 | 210 | 0.9 | 0.06 | 214 | 1.3 | 0.15 | 480 | 0.8 | 0.10 | 228 | 0.8 | 0.11 | 681 | 0.3 | 0.07 | 363 | 1.1 | 0.24 | 1233 | 0.4 | 0.14 |
| GFDL-CM3 | 248 | 1.7 | 0.15 | 355 | 1.2 | 0.13 | 192 | 2.3 | 0.22 | 1029 | 0.7 | 0.21 | 463 | 0.9 | 0.20 | 1373 | 0.3 | 0.14 | 217 | 1.2 | 0.15 | 1204 | 0.4 | 0.14 |
| GFDL-ESM2G | 168 | 2.4 | 0.12 | 311 | 1.9 | 0.15 | 96 | 2.9 | 0.12 | 1141 | 0.9 | 0.30 | 273 | 1.2 | 0.17 | 1340 | 0.4 | 0.16 | 214 | 1.8 | 0.19 | 1508 | 0.4 | 0.19 |
| GFDL-ESM2M | 156 | 2.4 | 0.12 | 307 | 1.8 | 0.14 | 118 | 2.7 | 0.15 | 1116 | 0.9 | 0.30 | 309 | 1.1 | 0.19 | 1441 | 0.4 | 0.17 | 200 | 1.8 | 0.17 | 1434 | 0.5 | 0.20 |
| HadGEM2-CC | 341 | 1.1 | 0.19 | 493 | 1.1 | 0.12 | 281 | 1.2 | 0.16 | 1058 | 1.1 | 0.28 | 469 | 0.8 | 0.22 | 1370 | 0.3 | 0.16 | 178 | 1.4 | 0.14 | 1158 | 0.6 | 0.22 |
| HadGEM2-ES | 332 | 1.1 | 0.19 | 469 | 1.1 | 0.12 | 259 | 1.1 | 0.15 | 1054 | 1.1 | 0.29 | 442 | 0.8 | 0.20 | 1283 | 0.4 | 0.16 | 180 | 1.4 | 0.14 | 1194 | 0.6 | 0.22 |
| inmcm4 | 266 | 1.5 | 0.18 | 251 | 1.1 | 0.08 | 275 | 1.5 | 0.25 | 1002 | 0.6 | 0.17 | 314 | 0.9 | 0.16 | 1246 | 0.3 | 0.10 | 552 | 0.7 | 0.19 | 1347 | 0.3 | 0.12 |
| IPSL-CM5A-LR | 218 | 2.3 | 0.17 | 111 | 2.3 | 0.06 | 176 | 2.5 | 0.21 | 438 | 1.5 | 0.16 | 233 | 1.3 | 0.18 | 746 | 0.7 | 0.17 | 294 | 1.3 | 0.22 | 1133 | 0.4 | 0.13 |
| IPSL-CM5A-MR | 165 | 2.4 | 0.15 | 103 | 2.2 | 0.05 | 144 | 2.7 | 0.19 | 542 | 1.3 | 0.17 | 316 | 1.4 | 0.21 | 1114 | 0.6 | 0.17 | 281 | 1.4 | 0.21 | 1119 | 0.4 | 0.14 |
| IPSL-CM5B-LR | 301 | 1.9 | 0.18 | 156 | 2.3 | 0.06 | 284 | 2.4 | 0.28 | 225 | 2.0 | 0.09 | 361 | 1.1 | 0.22 | 754 | 0.6 | 0.14 | 203 | 1.7 | 0.20 | 1257 | 0.6 | 0.25 |
| MIROC5 | 268 | 1.2 | 0.16 | 253 | 1.4 | 0.10 | 191 | 1.7 | 0.18 | 1572 | 0.9 | 0.42 | 426 | 1.2 | 0.30 | 1826 | 0.4 | 0.23 | 312 | 1.3 | 0.24 | 1298 | 0.4 | 0.17 |
| MIROC-ESM-CHEM | 318 | 1.4 | 0.17 | 665 | 0.9 | 0.16 | 243 | 1.8 | 0.21 | 1045 | 0.6 | 0.19 | 307 | 1.3 | 0.22 | 1215 | 0.4 | 0.15 | 305 | 1.3 | 0.20 | 1038 | 0.4 | 0.11 |
| MIROC-ESM | 332 | 1.4 | 0.19 | 682 | 0.8 | 0.16 | 233 | 1.9 | 0.21 | 1063 | 0.6 | 0.19 | 285 | 1.3 | 0.21 | 1204 | 0.4 | 0.14 | 307 | 1.3 | 0.20 | 1031 | 0.4 | 0.12 |
| MPI-ESM-LR | 220 | 2.3 | 0.16 | 254 | 1.8 | 0.09 | 146 | 2.9 | 0.18 | 1087 | 1.0 | 0.29 | 379 | 1.4 | 0.26 | 1902 | 0.5 | 0.27 | 208 | 1.9 | 0.22 | 1112 | 0.5 | 0.18 |
| MPI-ESM-MR | 251 | 2.2 | 0.17 | 256 | 1.9 | 0.11 | 153 | 2.9 | 0.18 | 1048 | 1.1 | 0.31 | 407 | 1.3 | 0.26 | 1922 | 0.5 | 0.32 | 174 | 2.0 | 0.18 | 1150 | 0.6 | 0.20 |
| MRI-CGCM3 | 180 | 1.9 | 0.12 | 182 | 1.8 | 0.06 | 141 | 2.3 | 0.16 | 333 | 1.8 | 0.14 | 317 | 1.3 | 0.21 | 1292 | 0.6 | 0.23 | 216 | 1.6 | 0.19 | 1100 | 0.7 | 0.24 |
| MRI-ESM1 | 189 | 1.8 | 0.12 | 175 | 1.9 | 0.06 | 155 | 2.2 | 0.17 | 344 | 1.9 | 0.15 | 334 | 1.2 | 0.21 | 1344 | 0.6 | 0.24 | 199 | 1.6 | 0.17 | 1103 | 0.7 | 0.25 |
| NorESM1-M | 249 | 1.7 | 0.16 | 409 | 1.3 | 0.15 | 251 | 1.7 | 0.23 | 1197 | 0.9 | 0.33 | 424 | 0.9 | 0.22 | 1984 | 0.4 | 0.26 | 292 | 0.9 | 0.16 | 1321 | 0.4 | 0.15 |

Table 4. Root mean Square Error (ε), Pattern Correlation (PC) and Standard Deviation (σ) for all seasonality indicators over the spatial domain for all considered time periods. Top three estimates closest (farthest) to observations are highlighted in yellow (red) while estimates within the observation uncertainty are marked green.

| Datasets | Annual P ε | PC | σ | RE ε | PC | σ | SI ε | PC | σ | Monsoon P ε | PC | σ | RE ε | PC | σ | SI ε | PC | σ | Westerly P ε | PC | σ | RE ε | PC | σ | SI ε | PC | σ |
|---|---|---|---|---|---|---|---|---|---|---|---|---|---|---|---|---|---|---|---|---|---|---|---|---|---|---|---|
| GPCP | 0 | 1.00 | 599 | 0.00 | 1.00 | 0.48 | 0.00 | 1.00 | 0.16 | 0 | 1.00 | 484 | 0.00 | 1.00 | 0.50 | 0.00 | 1.00 | 0.09 | 0 | 1.00 | 207 | 0.00 | 1.00 | 0.67 | 0.00 | 1.00 | 0.07 |
| CMAP | 428 | 0.87 | 809 | 0.13 | 0.98 | 0.53 | 0.10 | 0.83 | 0.19 | 315 | 0.85 | 587 | 0.08 | 0.99 | 0.51 | 0.06 | 0.84 | 0.10 | 127 | 0.95 | 297 | 0.22 | 0.96 | 0.72 | 0.04 | 0.93 | 0.09 |
| MMM | 489 | 0.78 | 647 | 0.40 | 0.66 | 0.45 | 0.12 | 0.75 | 0.17 | 411 | 0.78 | 552 | 0.35 | 0.73 | 0.43 | 0.08 | 0.65 | 0.06 | 170 | 0.84 | 253 | 0.61 | 0.67 | 0.68 | 0.12 | 0.79 | 0.12 |
| bcc-csm1-1 | 573 | 0.69 | 704 | 0.61 | 0.37 | 0.58 | 0.27 | 0.61 | 0.30 | 692 | 0.70 | 797 | 0.59 | 0.47 | 0.55 | 0.11 | 0.43 | 0.09 | 190 | 0.61 | 183 | 0.73 | 0.55 | 0.78 | 0.17 | 0.72 | 0.16 |
| BNU-ESM | 585 | 0.66 | 584 | 0.44 | 0.66 | 0.47 | 0.13 | 0.66 | 0.15 | 446 | 0.66 | 477 | 0.37 | 0.77 | 0.45 | 0.09 | 0.57 | 0.09 | 182 | 0.77 | 236 | 0.70 | 0.62 | 0.76 | 0.11 | 0.75 | 0.10 |
| CanESM2 | 538 | 0.79 | 811 | 0.40 | 0.75 | 0.48 | 0.13 | 0.67 | 0.13 | 342 | 0.79 | 535 | 0.38 | 0.77 | 0.43 | 0.10 | 0.47 | 0.06 | 261 | 0.89 | 389 | 0.61 | 0.67 | 0.68 | 0.11 | 0.73 | 0.12 |
| CCSM4 | 670 | 0.67 | 728 | 0.36 | 0.70 | 0.42 | 0.18 | 0.68 | 0.20 | 488 | 0.66 | 552 | 0.31 | 0.79 | 0.41 | 0.11 | 0.66 | 0.12 | 235 | 0.82 | 321 | 0.57 | 0.64 | 0.62 | 0.22 | 0.69 | 0.18 |
| CESM1-BGC | 650 | 0.67 | 720 | 0.35 | 0.71 | 0.42 | 0.18 | 0.70 | 0.20 | 479 | 0.67 | 554 | 0.30 | 0.80 | 0.41 | 0.11 | 0.67 | 0.12 | 218 | 0.82 | 306 | 0.57 | 0.65 | 0.63 | 0.21 | 0.71 | 0.18 |
| CESM1-CAM5 | 859 | 0.60 | 930 | 0.34 | 0.77 | 0.42 | 0.16 | 0.60 | 0.18 | 587 | 0.57 | 637 | 0.35 | 0.81 | 0.38 | 0.09 | 0.56 | 0.10 | 344 | 0.81 | 436 | 0.54 | 0.71 | 0.70 | 0.20 | 0.72 | 0.18 |
| CMCC-CESM | 650 | 0.79 | 882 | 0.52 | 0.62 | 0.63 | 0.14 | 0.69 | 0.18 | 538 | 0.75 | 721 | 0.55 | 0.73 | 0.78 | 0.08 | 0.59 | 0.08 | 187 | 0.74 | 243 | 0.71 | 0.58 | 0.81 | 0.14 | 0.73 | 0.13 |
| CMCC-CM | 618 | 0.73 | 822 | 0.54 | 0.69 | 0.68 | 0.20 | 0.72 | 0.24 | 506 | 0.73 | 686 | 0.52 | 0.77 | 0.74 | 0.11 | 0.68 | 0.13 | 193 | 0.67 | 233 | 0.77 | 0.67 | 0.93 | 0.22 | 0.72 | 0.18 |
| CMCC-CMS | 618 | 0.73 | 822 | 0.54 | 0.69 | 0.68 | 0.20 | 0.72 | 0.24 | 506 | 0.73 | 686 | 0.52 | 0.77 | 0.74 | 0.11 | 0.68 | 0.13 | 193 | 0.67 | 233 | 0.77 | 0.67 | 0.93 | 0.22 | 0.72 | 0.18 |
| CNRM-CM5 | 575 | 0.75 | 799 | 0.35 | 0.79 | 0.47 | 0.11 | 0.71 | 0.14 | 381 | 0.75 | 545 | 0.37 | 0.77 | 0.37 | 0.08 | 0.57 | 0.07 | 232 | 0.84 | 330 | 0.56 | 0.72 | 0.75 | 0.15 | 0.76 | 0.13 |
| CSIRO-Mk3-6-0 | 526 | 0.73 | 774 | 0.60 | 0.56 | 0.61 | 0.18 | 0.62 | 0.22 | 467 | 0.70 | 651 | 0.60 | 0.56 | 0.65 | 0.11 | 0.50 | 0.12 | 145 | 0.80 | 244 | 0.60 | 0.62 | 0.67 | 0.15 | 0.74 | 0.17 |
| EC-EARTH | 553 | 0.81 | 729 | 0.36 | 0.82 | 0.42 | 0.14 | 0.67 | 0.17 | 405 | 0.75 | 520 | 0.44 | 0.75 | 0.41 | 0.09 | 0.57 | 0.09 | 234 | 0.89 | 350 | 0.49 | 0.79 | 0.75 | 0.17 | 0.84 | 0.16 |
| FGOALS-g2 | 645 | 0.66 | 788 | 0.55 | 0.45 | 0.35 | 0.23 | 0.56 | 0.27 | 570 | 0.62 | 705 | 0.49 | 0.59 | 0.24 | 0.14 | 0.44 | 0.15 | 180 | 0.80 | 252 | 0.77 | 0.56 | 0.44 | 0.11 | 0.80 | 0.12 |
| GFDL-CM3 | 456 | 0.75 | 563 | 0.39 | 0.73 | 0.44 | 0.1 | 0.77 | 0.12 | 321 | 0.80 | 446 | 0.38 | 0.81 | 0.40 | 0.08 | 0.56 | 0.06 | 175 | 0.74 | 235 | 0.59 | 0.70 | 0.81 | 0.15 | 0.73 | 0.13 |
| GFDL-ESM2G | 503 | 0.77 | 699 | 0.49 | 0.64 | 0.64 | 0.11 | 0.79 | 0.16 | 341 | 0.82 | 521 | 0.42 | 0.76 | 0.64 | 0.07 | 0.64 | 0.08 | 212 | 0.79 | 321 | 0.84 | 0.66 | 1.07 | 0.15 | 0.72 | 0.15 |
| GFDL-ESM2M | 495 | 0.74 | 646 | 0.51 | 0.63 | 0.65 | 0.12 | 0.76 | 0.16 | 351 | 0.77 | 486 | 0.40 | 0.76 | 0.62 | 0.07 | 0.66 | 0.08 | 187 | 0.79 | 287 | 0.81 | 0.65 | 1.04 | 0.14 | 0.67 | 0.13 |
| HadGEM2-CC | 673 | 0.66 | 853 | 0.59 | 0.45 | 0.61 | 0.20 | 0.67 | 0.25 | 526 | 0.65 | 679 | 0.61 | 0.39 | 0.59 | 0.13 | 0.64 | 0.16 | 226 | 0.78 | 319 | 0.66 | 0.55 | 0.71 | 0.20 | 0.70 | 0.20 |
| HadGEM2-ES | 684 | 0.68 | 875 | 0.58 | 0.46 | 0.60 | 0.21 | 0.67 | 0.26 | 537 | 0.67 | 700 | 0.59 | 0.41 | 0.57 | 0.13 | 0.64 | 0.16 | 225 | 0.78 | 321 | 0.66 | 0.56 | 0.72 | 0.20 | 0.70 | 0.20 |
| inmcm4 | 592 | 0.73 | 706 | 0.53 | 0.71 | 0.40 | 0.14 | 0.64 | 0.10 | 382 | 0.72 | 480 | 0.46 | 0.76 | 0.35 | 0.10 | 0.55 | 0.05 | 288 | 0.82 | 341 | 0.76 | 0.67 | 0.53 | 0.12 | 0.51 | 0.07 |
| IPSL-CM5A-LR | 467 | 0.76 | 712 | 0.62 | 0.58 | 0.76 | 0.13 | 0.70 | 0.13 | 344 | 0.77 | 494 | 0.59 | 0.61 | 0.74 | 0.09 | 0.54 | 0.07 | 209 | 0.81 | 316 | 0.72 | 0.65 | 0.94 | 0.14 | 0.68 | 0.13 |
| IPSL-CM5A-MR | 468 | 0.77 | 727 | 0.60 | 0.55 | 0.72 | 0.11 | 0.76 | 0.14 | 338 | 0.78 | 524 | 0.55 | 0.62 | 0.70 | 0.09 | 0.59 | 0.07 | 195 | 0.82 | 305 | 0.72 | 0.62 | 0.90 | 0.14 | 0.74 | 0.13 |
| IPSL-CM5B-LR | 536 | 0.64 | 641 | 0.76 | 0.45 | 0.81 | 0.17 | 0.54 | 0.19 | 450 | 0.64 | 530 | 0.79 | 0.45 | 0.82 | 0.11 | 0.38 | 0.11 | 206 | 0.50 | 205 | 0.74 | 0.56 | 0.87 | 0.17 | 0.52 | 0.16 |
| MIROC5 | 672 | 0.80 | 803 | 0.38 | 0.65 | 0.39 | 0.23 | 0.74 | 0.26 | 551 | 0.76 | 692 | 0.34 | 0.76 | 0.42 | 0.13 | 0.70 | 0.15 | 239 | 0.78 | 316 | 0.60 | 0.57 | 0.53 | 0.25 | 0.74 | 0.23 |
| MIROC-ESM-CHEM | 797 | 0.54 | 739 | 0.50 | 0.55 | 0.40 | 0.14 | 0.61 | 0.15 | 599 | 0.48 | 519 | 0.51 | 0.65 | 0.31 | 0.08 | 0.56 | 0.07 | 285 | 0.76 | 369 | 0.63 | 0.63 | 0.72 | 0.15 | 0.68 | 0.13 |
| MIROC-ESM | 797 | 0.54 | 739 | 0.50 | 0.55 | 0.40 | 0.14 | 0.61 | 0.15 | 599 | 0.48 | 519 | 0.51 | 0.65 | 0.31 | 0.08 | 0.56 | 0.07 | 285 | 0.76 | 369 | 0.63 | 0.63 | 0.72 | 0.15 | 0.68 | 0.13 |
| MPI-ESM-LR | 621 | 0.71 | 802 | 0.48 | 0.74 | 0.67 | 0.18 | 0.70 | 0.21 | 478 | 0.71 | 628 | 0.43 | 0.80 | 0.69 | 0.10 | 0.65 | 0.12 | 177 | 0.79 | 270 | 0.82 | 0.71 | 1.01 | 0.21 | 0.76 | 0.19 |
| MPI-ESM-MR | 654 | 0.72 | 821 | 0.48 | 0.72 | 0.61 | 0.22 | 0.71 | 0.24 | 541 | 0.72 | 683 | 0.43 | 0.78 | 0.65 | 0.12 | 0.67 | 0.13 | 171 | 0.72 | 235 | 0.82 | 0.70 | 0.98 | 0.20 | 0.73 | 0.18 |
| MRI-CGCM3 | 539 | 0.66 | 637 | 0.72 | 0.70 | 0.76 | 0.16 | 0.59 | 0.19 | 463 | 0.62 | 509 | 0.66 | 0.62 | 0.67 | 0.11 | 0.47 | 0.11 | 150 | 0.74 | 202 | 0.81 | 0.68 | 0.94 | 0.14 | 0.74 | 0.15 |
| MRI-ESM1 | 527 | 0.66 | 619 | 0.75 | 0.71 | 0.80 | 0.16 | 0.62 | 0.19 | 453 | 0.63 | 504 | 0.69 | 0.64 | 0.71 | 0.11 | 0.51 | 0.12 | 159 | 0.69 | 187 | 0.82 | 0.66 | 0.93 | 0.13 | 0.75 | 0.14 |
| NorESM1-M | 582 | 0.64 | 609 | 0.39 | 0.66 | 0.43 | 0.15 | 0.64 | 0.16 | 448 | 0.65 | 490 | 0.33 | 0.79 | 0.47 | 0.09 | 0.60 | 0.10 | 187 | 0.76 | 258 | 0.62 | 0.60 | 0.57 | 0.18 | 0.63 | 0.16 |

Table 5. Percentage change in seasonality indicators for future period (2061-2100) for WPR and MPR for all basins under RCP8.5 scenario relative to the historical period (1961-2000). Note: Negative values suggest decrease in P, SI and RE while positive values suggest the opposite.

| Basins→ | Indus | | | | | | Ganges | | | | | | Brahmaputra | | | | | | Mekong | | | | | |
|---|---|---|---|---|---|---|---|---|---|---|---|---|---|---|---|---|---|---|---|---|---|---|---|---|
| | Westerly | | | Monsoon | | | Westerly | | | Monsoon | | | Westerly | | | Monsoon | | | Westerly | | | Monsoon | | |
| Models | P% | RE% | SI% | P% | RE% | SI% | P% | RE% | SI% | P% | RE% | SI% | P% | RE% | SI% | P% | RE% | SI% | P% | RE% | SI% | P% | RE% | SI% |
| bcc-csm1-1 | 7 | 6 | 20 | 3 | 0 | 183 | 14 | -1 | 19 | 14 | -10 | 197 | 14 | -1 | 15 | -3 | 10 | 191 | 1 | 3 | 6 | -15 | 13 | 153 |
| BNU-ESM | -1 | 5 | 3 | 41 | -6 | 29 | -4 | 8 | 1 | 19 | 1 | 20 | 0 | 7 | 6 | 11 | 9 | 16 | 3 | 17 | 15 | 9 | 19 | 17 |
| CanESM2 | 22 | -8 | 18 | 5 | -11 | -1 | 26 | -17 | 5 | 20 | 25 | 51 | 41 | -5 | 37 | 47 | 38 | 99 | 4 | 20 | 28 | 12 | 27 | 47 |
| CCSM4 | 1 | 6 | 14 | 9 | 9 | 20 | -7 | 1 | -4 | 16 | 8 | 32 | 3 | 10 | 9 | 16 | 16 | 34 | 10 | 25 | 36 | 13 | 19 | 23 |
| CESM1-BGC | 1 | 6 | 14 | 13 | 12 | 27 | -6 | -1 | -7 | 15 | 10 | 27 | 2 | 15 | 14 | 19 | 15 | 39 | 4 | 12 | 19 | 14 | 28 | 38 |
| CESM1-CAM5 | 17 | 2 | 22 | 9 | 1 | 17 | 18 | -9 | 6 | 13 | 18 | 25 | 36 | 4 | 35 | 26 | 23 | 48 | 10 | 14 | 25 | 12 | 10 | 29 |
| CMCC-CESM | -8 | 15 | 7 | -33 | 7 | -21 | 1 | 6 | 8 | -4 | 8 | -1 | -16 | 13 | -2 | 2 | 1 | 14 | -4 | 16 | 7 | 10 | 17 | 25 |
| CMCC-CM | -4 | 11 | 0 | 12 | 4 | 31 | -24 | 12 | -10 | 29 | 7 | 41 | -6 | 27 | 22 | 19 | 14 | 48 | -4 | 17 | 6 | 20 | 25 | 37 |
| CMCC-CMS | -18 | 18 | -4 | 5 | 4 | 9 | -30 | 14 | -19 | 3 | 13 | 13 | -2 | 33 | 28 | 10 | 27 | 37 | -8 | 16 | 2 | 22 | 5 | 32 |
| CNRM-CM5 | 10 | 0 | 6 | 15 | -12 | 2 | 8 | 0 | 5 | 18 | -7 | 14 | 27 | 1 | 25 | 23 | -13 | 20 | -1 | 8 | 2 | 9 | 7 | 21 |
| CSIRO-Mk3-6-0 | 28 | 1 | 21 | 31 | 1 | 60 | 15 | -1 | 13 | 31 | -12 | 34 | 29 | 0 | 35 | 4 | 14 | 17 | -17 | 15 | -3 | 7 | 7 | 12 |
| EC-EARTH | 3 | 15 | 10 | 0 | 20 | 22 | -12 | 8 | -8 | 11 | 38 | 41 | 11 | 9 | 27 | 16 | 18 | 62 | -2 | 12 | 7 | 10 | 37 | 36 |
| FGOALS-g2 | -16 | 24 | -1 | 27 | -12 | 4 | -16 | 16 | 2 | 47 | -27 | 23 | 1 | 25 | 23 | 16 | -2 | 8 | 43 | -8 | 16 | -16 | -7 | -19 |
| GFDL-CM3 | -8 | 12 | 15 | 31 | -22 | 1 | -22 | 2 | -18 | 28 | -13 | 19 | 8 | -3 | 7 | 29 | 15 | 41 | 8 | 8 | 1 | 12 | 4 | 29 |
| GFDL-ESM2G | 1 | 6 | 26 | 1 | 2 | -4 | 22 | -1 | 38 | 8 | 20 | 22 | 21 | -7 | 17 | 25 | 19 | 62 | 24 | -1 | 25 | 8 | 7 | 17 |
| GFDL-ESM2M | 14 | 4 | 30 | -6 | 11 | -10 | 11 | 2 | 19 | 7 | 34 | 27 | 4 | 0 | 0 | 20 | 19 | 64 | 1 | 6 | 11 | 4 | 13 | 35 |
| HadGEM2-CC | 0 | 19 | 21 | 25 | 0 | 33 | -7 | 4 | 18 | 35 | -7 | 42 | -8 | 29 | 23 | 7 | 45 | 34 | 14 | 13 | 34 | -6 | 28 | 19 |
| HadGEM2-ES | -10 | 21 | 6 | 18 | 5 | 26 | -13 | 20 | 15 | 27 | -3 | 31 | -10 | 36 | 34 | 10 | 9 | 29 | 23 | 8 | 42 | -7 | 26 | 19 |
| inmcm4 | -13 | 7 | -6 | 19 | 0 | 15 | -20 | 6 | -20 | 14 | 3 | 10 | 3 | 1 | 5 | 10 | -15 | 4 | 15 | 15 | 18 | 15 | -6 | 11 |
| IPSL-CM5A-LR | -15 | 6 | 4 | 6 | -2 | 14 | -32 | 14 | -19 | 34 | -5 | 39 | -32 | 29 | -19 | 18 | -5 | 16 | -13 | 24 | -3 | 4 | 17 | 32 |
| IPSL-CM5A-MR | -8 | 8 | 16 | 48 | -7 | 56 | -21 | 9 | -7 | 33 | -7 | 32 | -44 | 45 | -13 | 8 | 6 | 29 | -11 | 24 | 3 | 10 | 23 | 32 |
| IPSL-CM5B-LR | 8 | 11 | 21 | 19 | -9 | 30 | -13 | 6 | -10 | 69 | -18 | 67 | -11 | 10 | 1 | 25 | -5 | 17 | 1 | 4 | 0 | -2 | 6 | -1 |
| MIROC5 | -1 | 22 | 11 | 67 | -9 | 43 | 3 | 10 | 18 | 15 | -8 | 5 | 40 | 7 | 49 | 14 | 7 | 26 | 34 | -13 | 13 | 11 | -8 | 3 |
| MIROC-ESM-CHEM | -16 | 13 | -5 | 9 | -6 | 6 | -19 | 19 | -1 | 17 | 3 | 24 | 12 | -10 | 1 | 16 | 4 | 28 | 3 | -3 | 0 | 1 | -12 | 3 |
| MIROC-ESM | -20 | 3 | -19 | 12 | 0 | 5 | -18 | 0 | -14 | 15 | 3 | 25 | 25 | -10 | 12 | 15 | 1 | 31 | 5 | -6 | 1 | 4 | -15 | -10 |
| MPI-ESM-LR | -13 | 14 | 0 | -4 | 12 | 1 | -25 | 14 | -10 | -1 | 7 | 5 | -16 | 25 | 1 | 3 | 9 | 26 | -12 | 20 | -5 | 19 | 20 | 41 |
| MPI-ESM-MR | -4 | 12 | 3 | -2 | 11 | 2 | -9 | 6 | 6 | -4 | 9 | 4 | -2 | 15 | 11 | 0 | 20 | 13 | -1 | 10 | 13 | 15 | 9 | 34 |
| MRI-CGCM3 | 26 | -6 | 20 | 16 | 1 | 13 | 19 | -8 | 12 | 28 | 0 | 35 | 34 | -6 | 30 | 17 | 7 | 32 | 0 | 6 | 9 | 16 | 13 | 35 |
| MRI-ESM1 | 26 | 0 | 30 | 22 | -3 | 18 | 13 | -2 | 17 | 31 | -5 | 39 | 26 | 7 | 35 | 9 | 9 | 23 | 15 | 8 | 36 | 18 | 14 | 33 |
| NorESM1-M | -9 | 6 | -2 | 23 | -5 | 24 | -3 | 12 | 7 | 17 | 1 | 17 | 1 | 13 | 13 | 12 | 17 | 24 | 39 | 7 | 47 | 16 | -6 | 23 |

Table 6. Estimates for the basin integrated seasonality indicators, Precipitation (P), Relative Entropy (RE) and Seasonality Index (SI) for the historical period (1961-2000). Note: negative values imply a increase (decrease) in number of dry (wet) days, whereas positive values imply decrease (increase) in number of dry (wet) days.

| Basins→ | Indus | | Ganges | | Brahma | | Mekong | |
|---|---|---|---|---|---|---|---|---|
| Models | WPR RE (days) | MPR RE (days) | WPR RE (days) | MPR RE (days) | WPR RE (days) | MPR RE (days) | WPR RE (days) | MPR RE (days) |
| bcc-csm1-1 | -4 | 0 | 1 | 7 | 1 | -5 | -2 | -7 |
| BNU-ESM | -3 | 4 | -5 | -1 | -4 | -4 | -10 | -6 |
| CanESM2 | 6 | 8 | 12 | -14 | 3 | -16 | -12 | -8 |
| CCSM4 | -4 | -6 | 0 | -5 | -6 | -7 | -14 | -7 |
| CESM1-BGC | -4 | -7 | 1 | -6 | -9 | -7 | -7 | -11 |
| CESM1-CAM5 | -1 | -1 | 6 | -9 | -3 | -9 | -9 | -4 |
| CMCC-CESM | -8 | -4 | -4 | -5 | -8 | -1 | -10 | -5 |
| CMCC-CM | -6 | -2 | -5 | -5 | -16 | -8 | -9 | -12 |
| CMCC-CMS | -10 | -2 | -7 | -8 | -19 | -12 | -9 | -3 |
| CNRM-CM5 | 0 | 8 | 0 | 4 | -1 | 6 | -5 | -3 |
| CSIRO-Mk3-6-0 | -1 | 0 | 1 | 9 | 0 | -6 | -9 | -3 |
| EC-EARTH | -9 | -11 | -5 | -14 | -6 | -5 | -7 | -11 |
| FGOALS-g2 | -14 | 7 | -10 | 17 | -14 | 1 | 5 | 3 |
| GFDL-CM3 | -8 | 16 | -1 | 8 | 2 | -5 | -5 | -1 |
| GFDL-ESM2G | -3 | -1 | 1 | -12 | 5 | -7 | 1 | -3 |
| GFDL-ESM2M | -2 | -7 | -1 | -19 | 0 | -7 | -4 | -6 |
| HadGEM2-CC | -11 | 0 | -2 | 5 | -16 | -13 | -8 | -13 |
| HadGEM2-ES | -13 | -3 | -12 | 2 | -19 | -3 | -5 | -13 |
| inmcm4 | -4 | 0 | -4 | -1 | -1 | 5 | -8 | 2 |
| IPSL-CM5A-LR | -4 | 1 | -7 | 3 | -17 | 3 | -15 | -7 |
| IPSL-CM5A-MR | -5 | 5 | -4 | 5 | -25 | -3 | -14 | -9 |
| IPSL-CM5B-LR | -7 | 6 | -3 | 13 | -6 | 3 | -3 | -3 |
| MIROC5 | -13 | 6 | -6 | 5 | -5 | -3 | 9 | 3 |
| MIROC-ESM-CHEM | -8 | 4 | -11 | -2 | 7 | -1 | 2 | 5 |
| MIROC-ESM | -2 | 0 | 0 | -2 | 7 | 0 | 4 | 6 |
| MPI-ESM-LR | -7 | -7 | -6 | -4 | -15 | -4 | -11 | -9 |
| MPI-ESM-MR | -7 | -7 | -3 | -6 | -9 | -9 | -6 | -5 |
| MRI-CGCM3 | 4 | 0 | 5 | 0 | 4 | -4 | -4 | -7 |
| MRI-ESM1 | 0 | 2 | 1 | 3 | -5 | -4 | -5 | -7 |
| NorESM1-M | -4 | 3 | -8 | -1 | -8 | -6 | -4 | 2 |

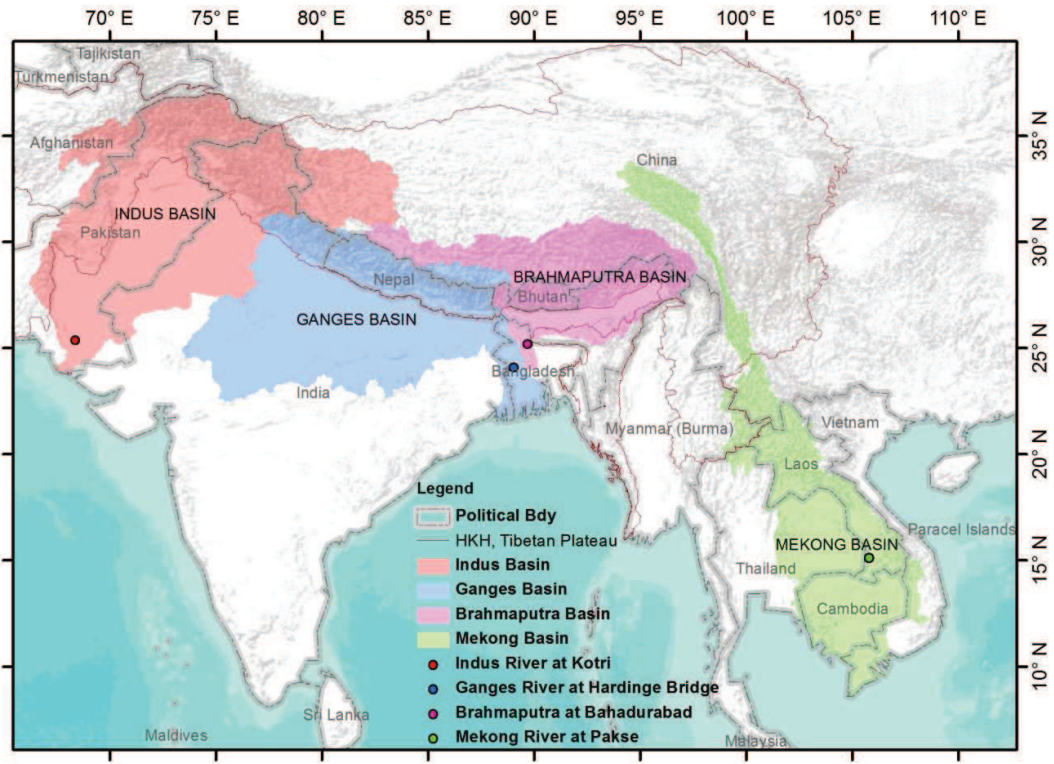

Figure 1. Study Area (left to right), Indus, Ganges, Brahmaputra and Mekong basins

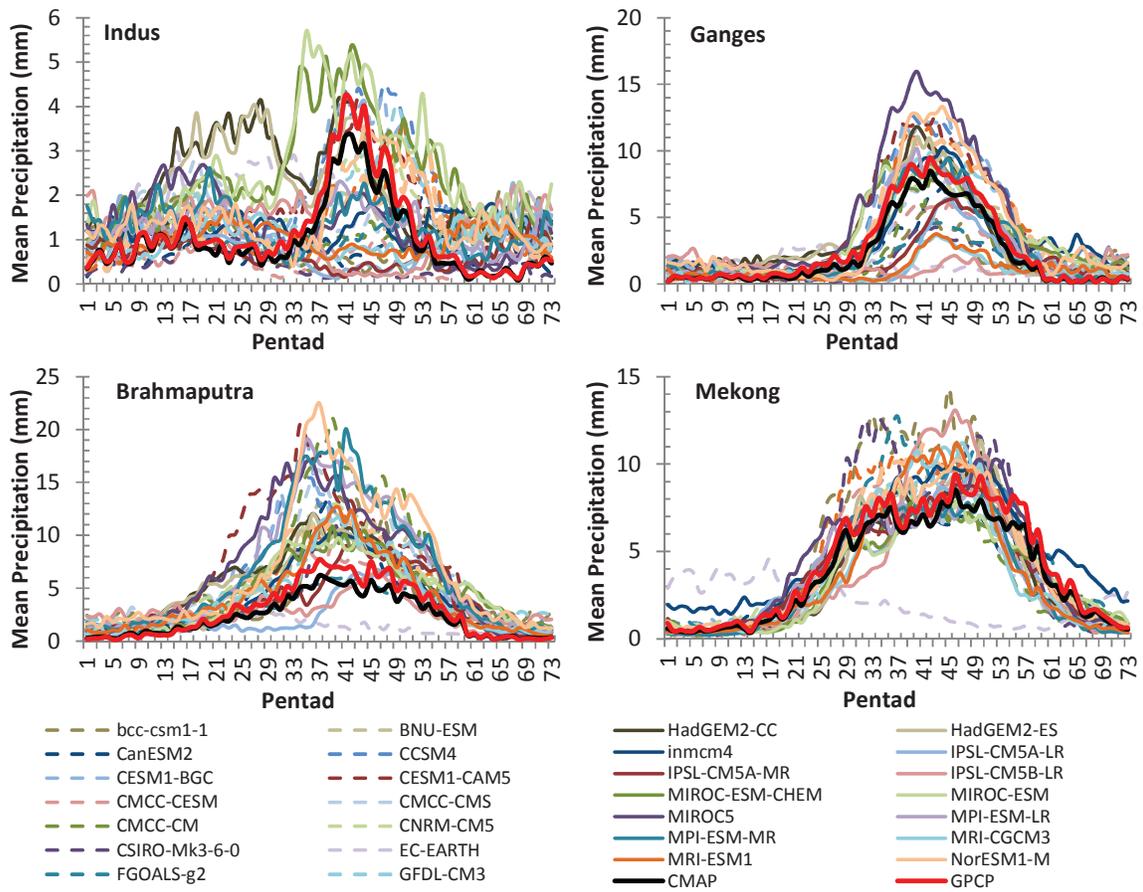

**Figure 2:** Climatological mean basin-integrated pentad precipitation (mm) for the Indus, Ganges, Brahmaputra and Mekong basins for the CMIP5 climate models and for the GPCP/CMAP observations

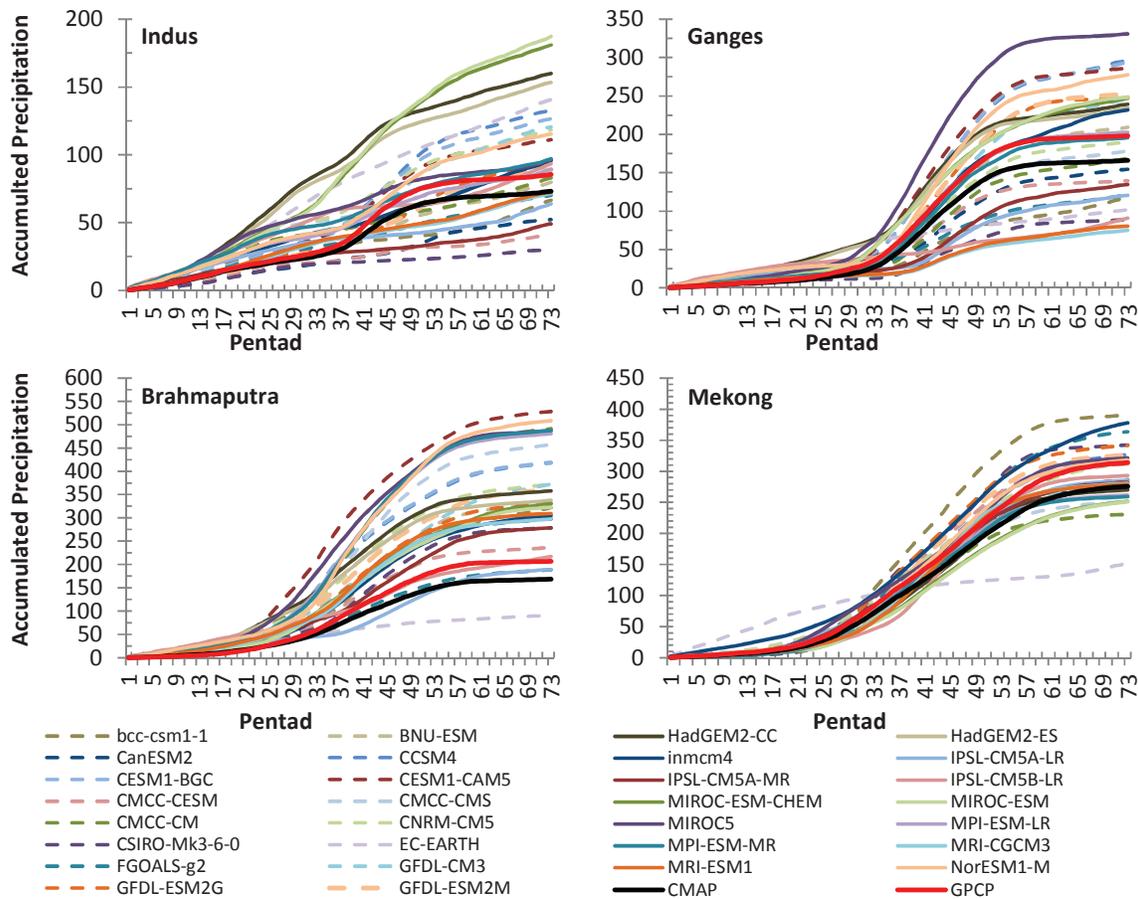

**Figure 3:** Same as figure 2 but for the accumulated basin-integrated mean pentad precipitation (mm)

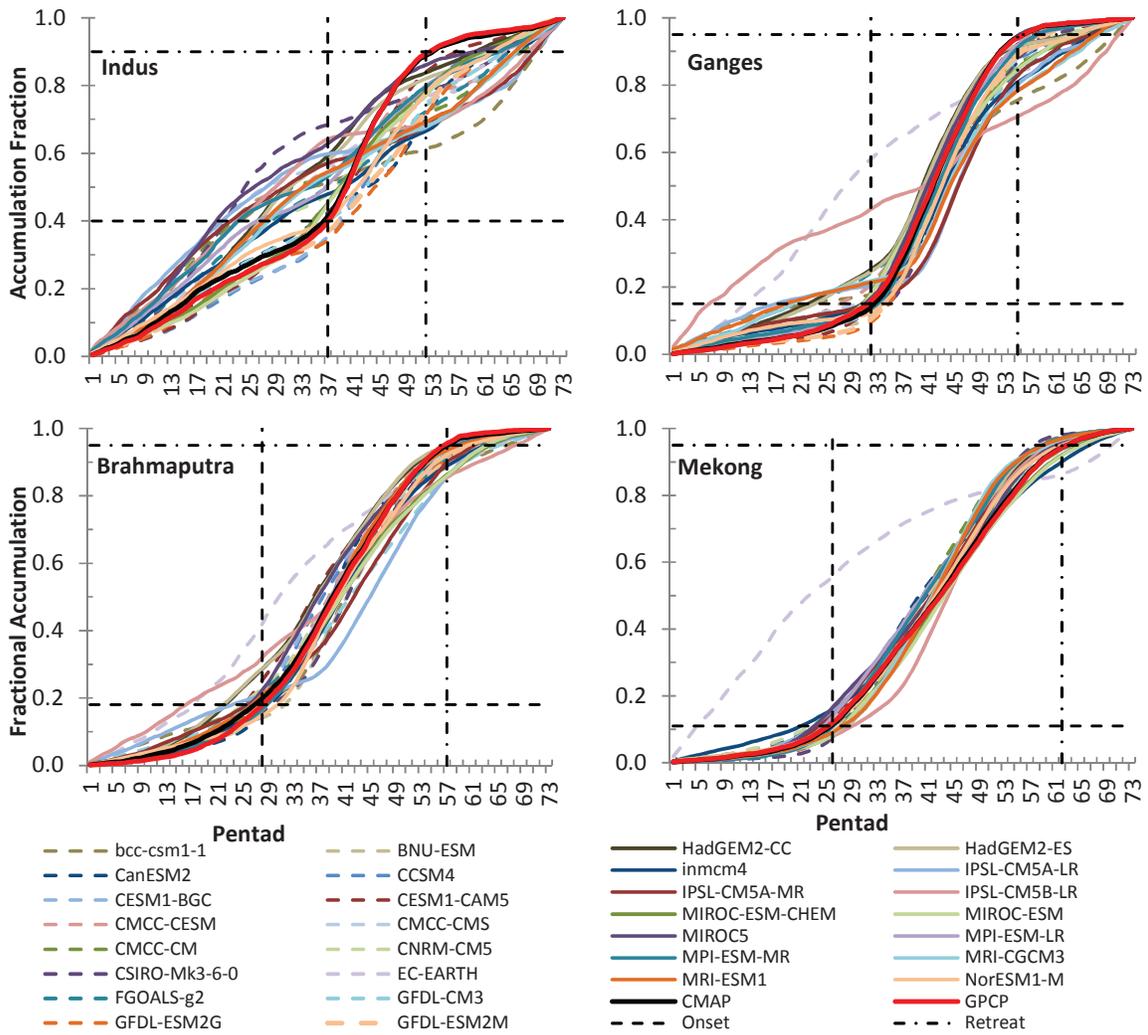

**Figure 4:** Same as figure 2 but for the fractional accumulated basin-integrated mean pentad precipitation. Dotted lines show timings of the monsoon onset and retreat for each basin.

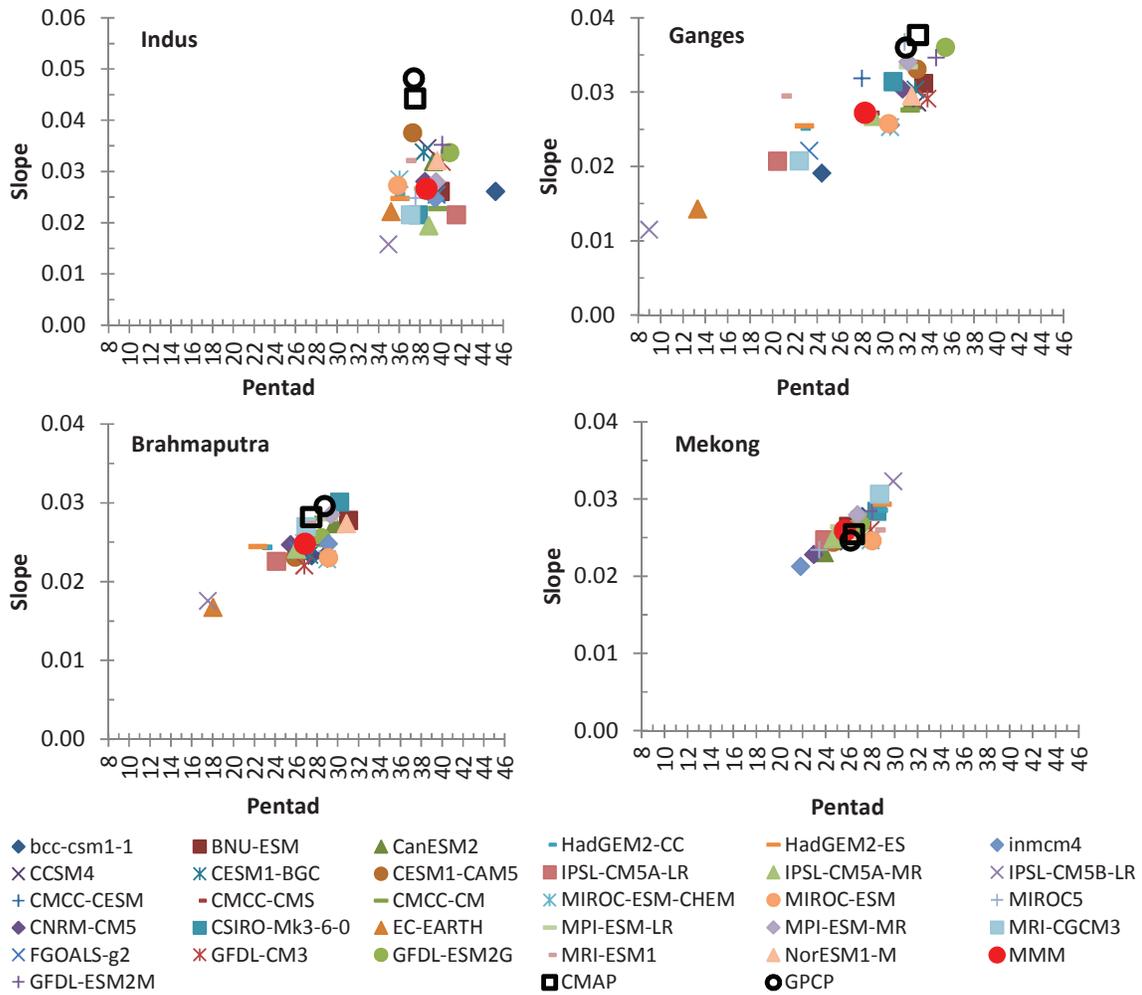

Figure 5. The RFA slope for the active monsoonal duration (onset to retreat) plotted against the monsoon onset pentad.

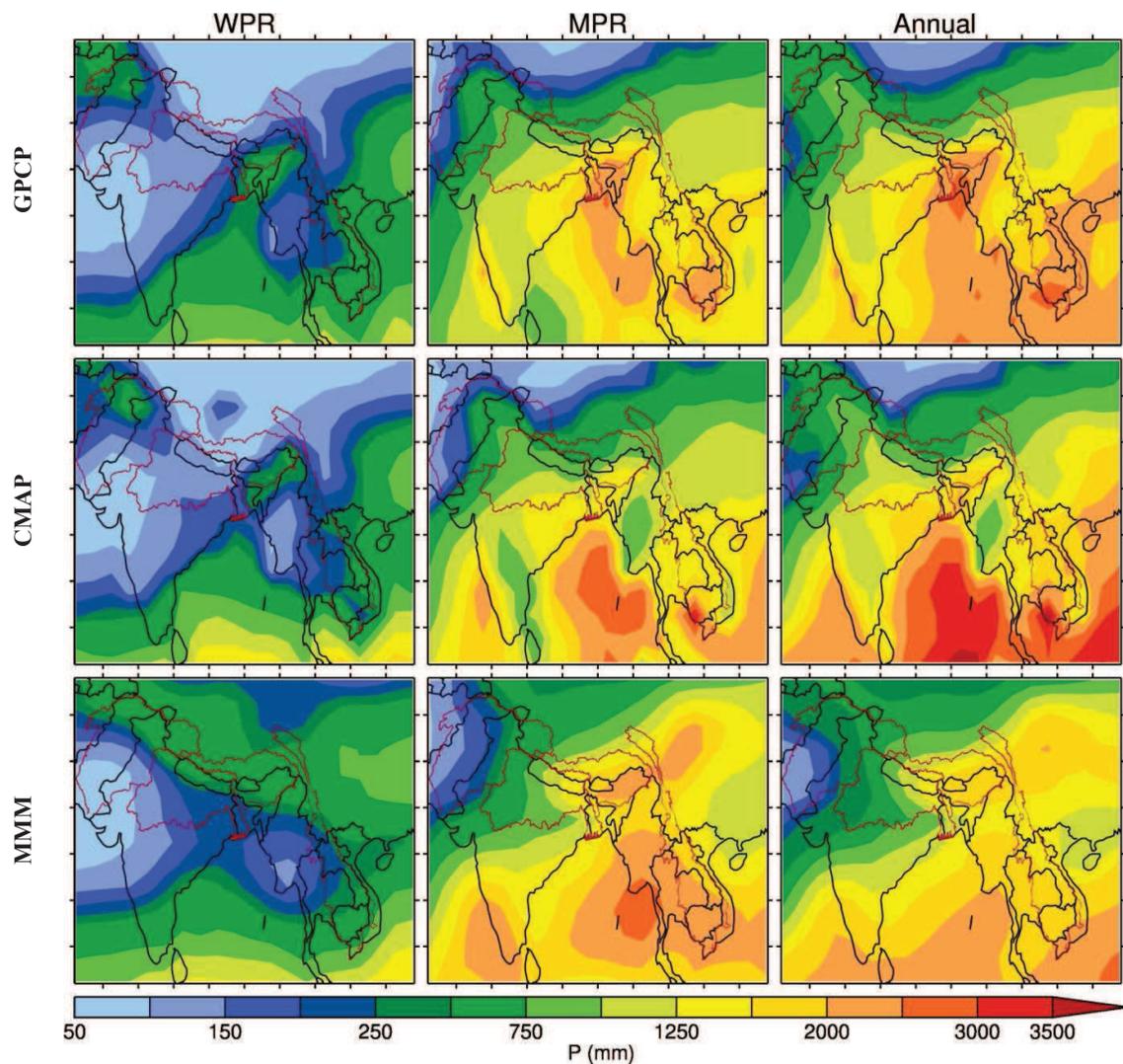

**Figure 6. Mean total precipitation in mm from GPCP, CMAP and MMM dataset shown in top, middle and bottom row, respectively for the WPR, MPR and annual precipitation regimes in the left, middle and right columns, respectively. Note: spatial biases in P from individual model with respect to GPCP are given in the supplement. Political boundaries are shown in black and study basins are shown in red.**

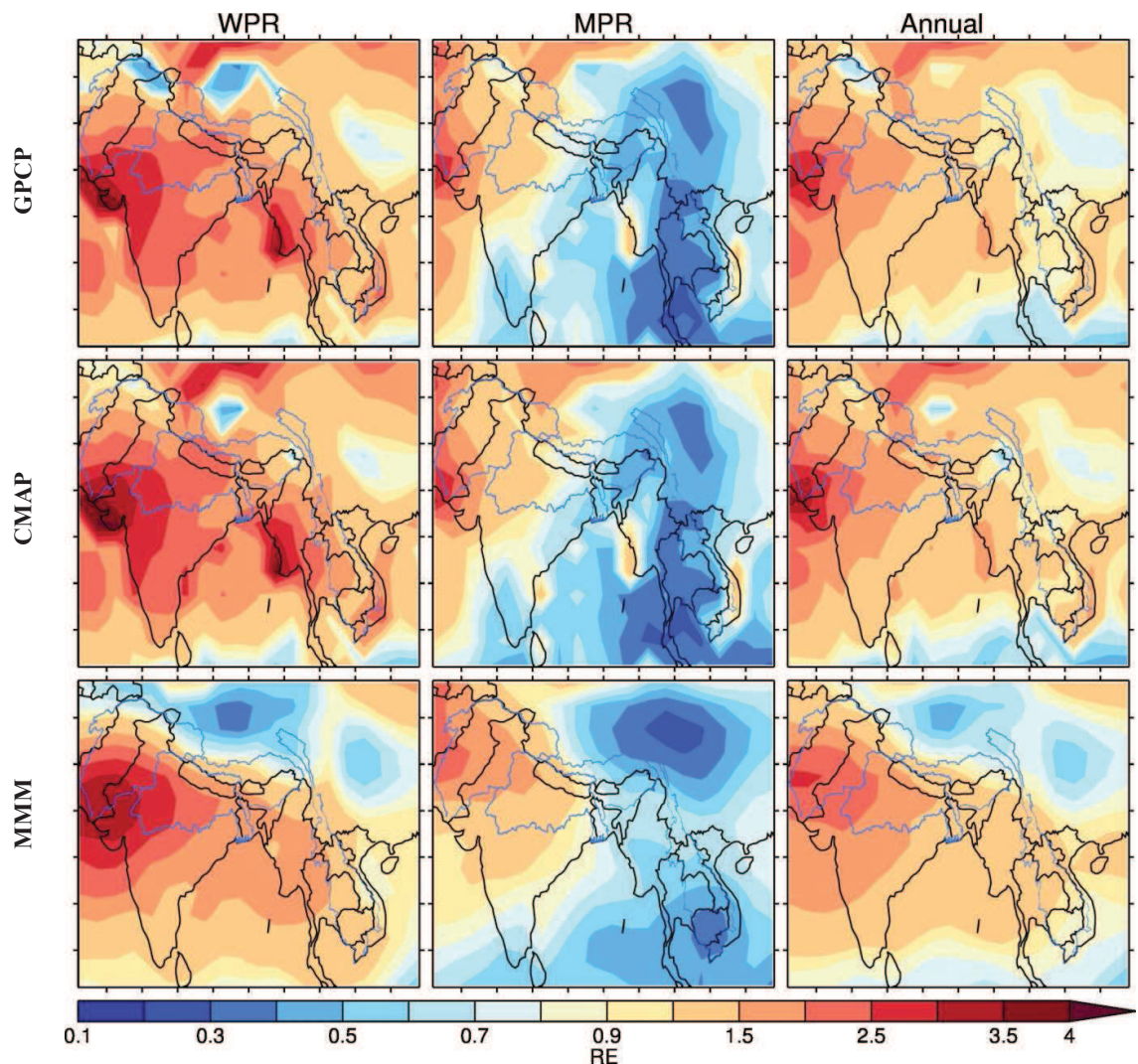
Figure 7. Same as Figure 6, but for RE. Study basins are shown in blue here.

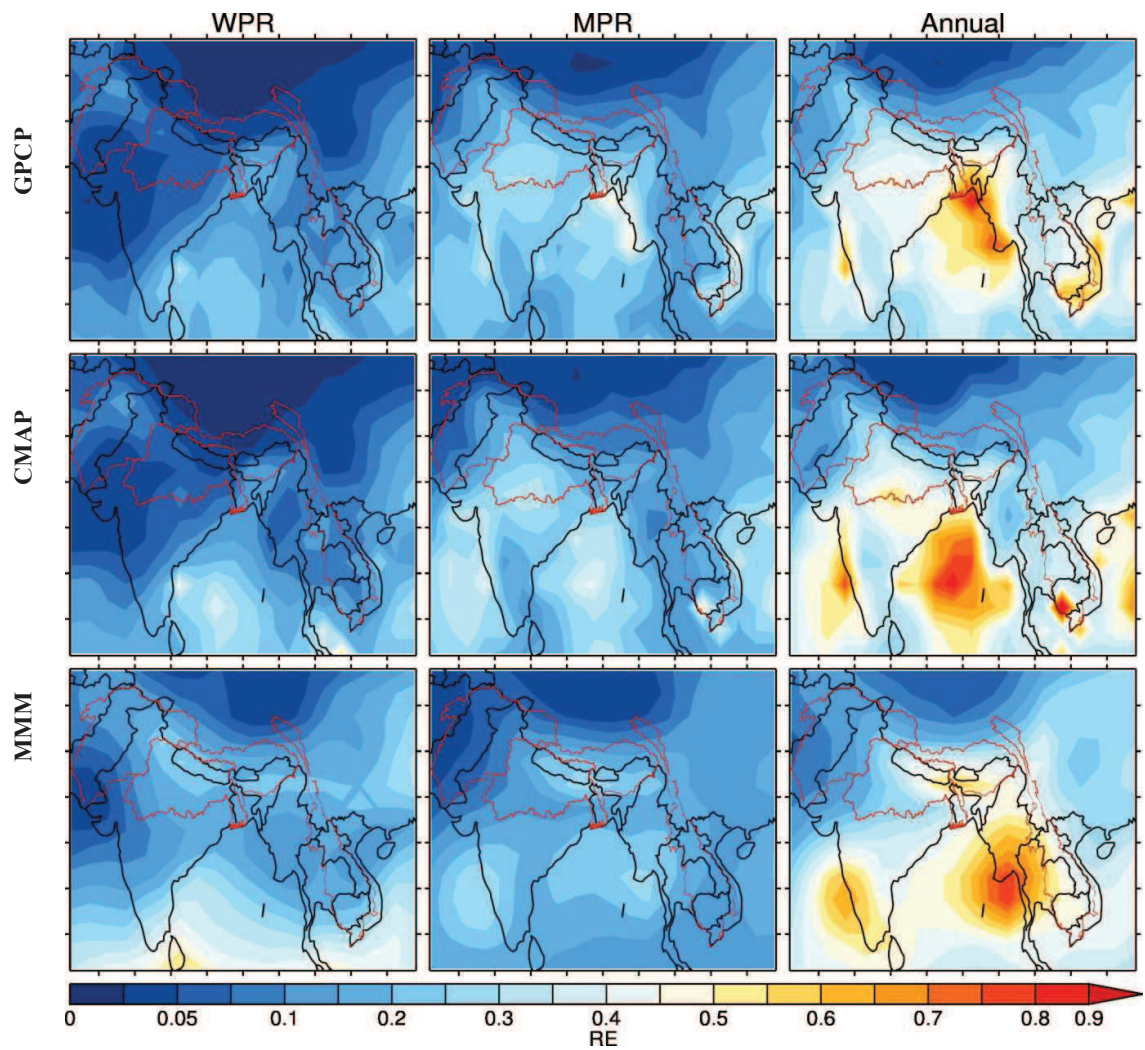

**Figure 8.** Same as Figure 6, but for SI

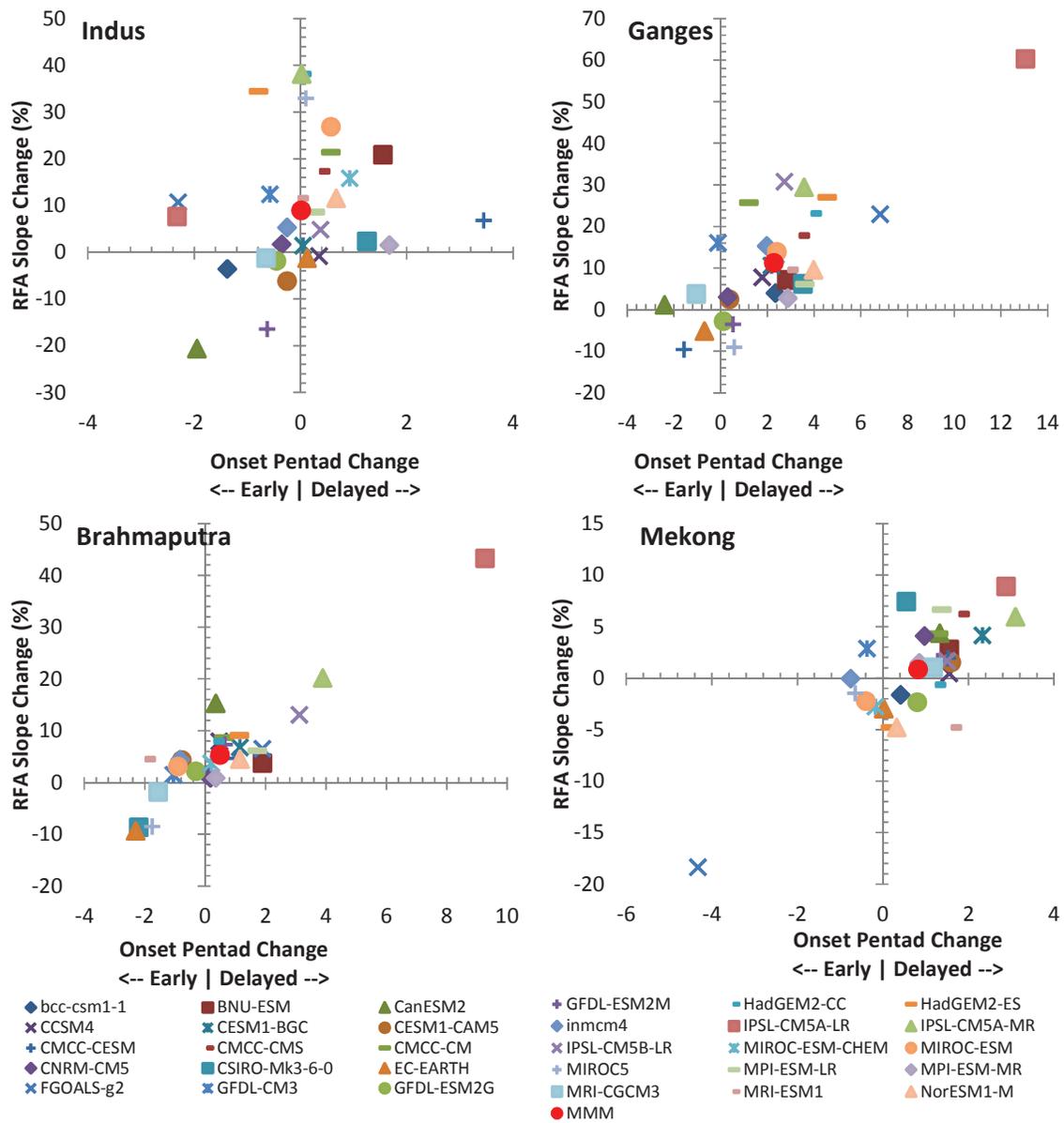

**Figure 9.** Future changes in the timings of monsoon onset (in pentads) and RFA slope (%) for the period (2061-2100) under the RCP8.5 scenario relative to the historical period (1961-2000). For statistically significant changes in RFA slope and onset see Table 2.

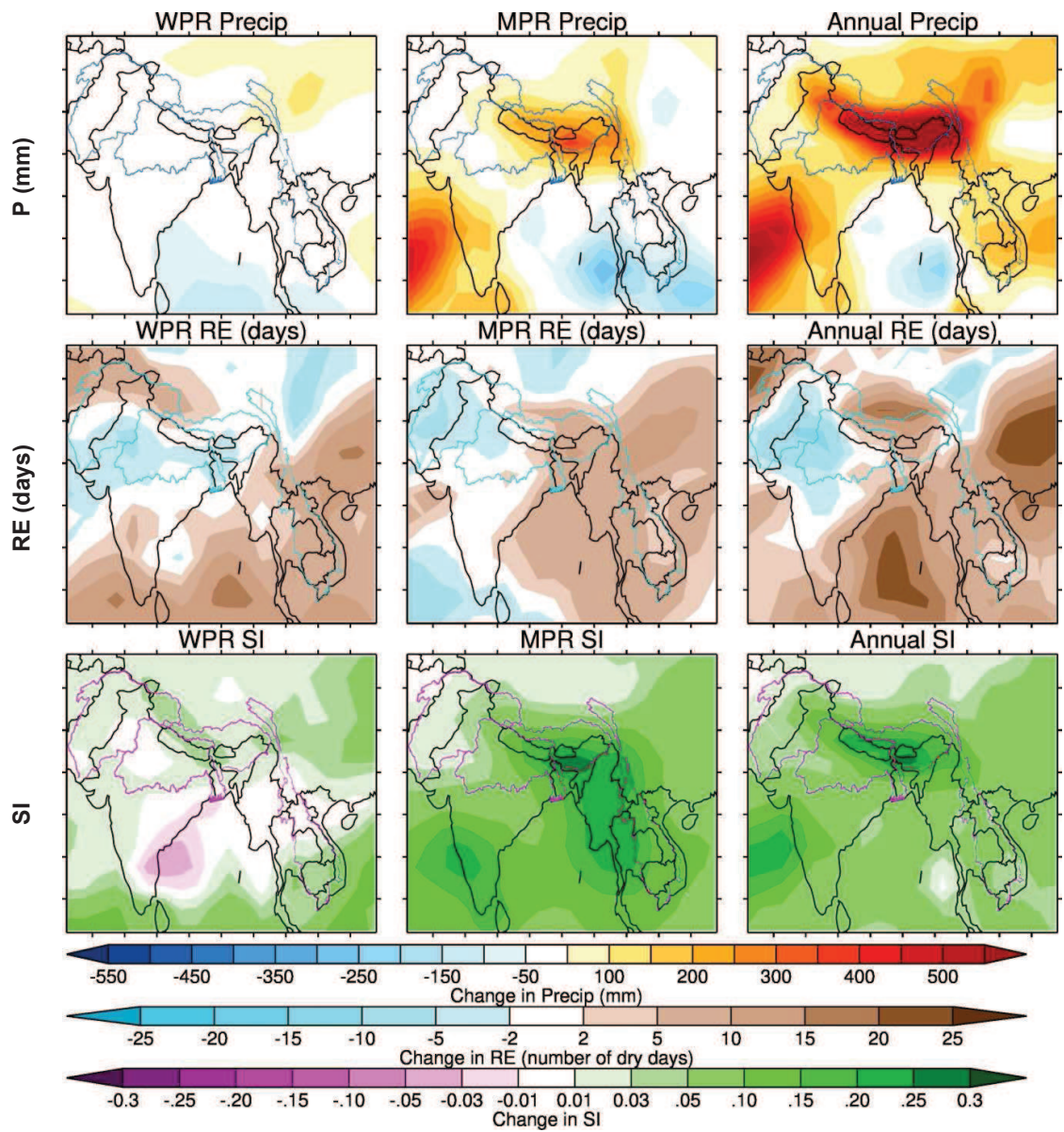

Figure 10. The MMM change for the future period (2061-2100) in the seasonality indicators, P, RE, SI for the WPR, MPR and annual precipitation regime under the RCP8.5 scenario with respect to historical period (1961-2000). Note: change in RE is interpreted in terms of increase (positive values) or decrease (negative values) in number of dry days. Spatial scale changes in P, RE and SI from the individual models are given in the supplement.

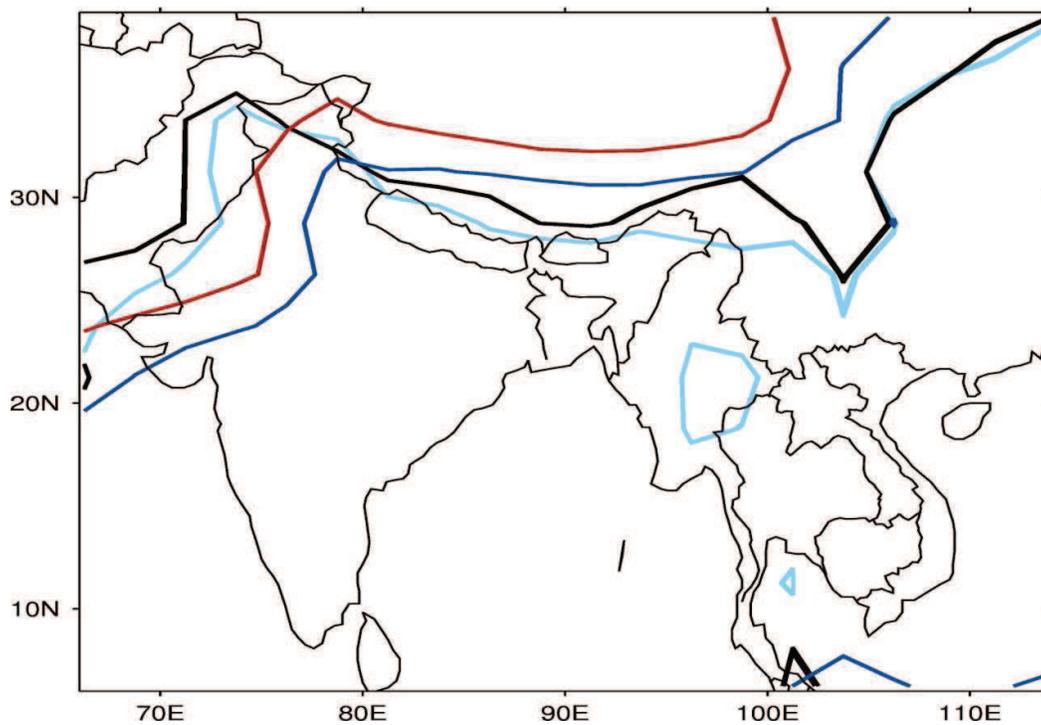

**Figure 11.** Spatial extent of the monsoon as estimated by SI=0.11 from the GPCP (black), the CMAP (cyan), MMM historical (blue) and MMM RCP85 (red) lines.